%% file: memo_signals_ramo.tex
\documentclass[10pt]{article}
\usepackage[twoside,pdftex,papersize={210mm,280mm}, total={16.8cm,23.6cm}, left=1.5cm, top=1.5cm, right=1.5cm, bottom=2.5cm, includehead=true]{geometry}
\usepackage{extarrows}

\usepackage{amstext}
\usepackage{amsmath}
\usepackage{amssymb}
\usepackage{amsfonts}
\usepackage{graphicx}  % to include figures (can also use other packages)
\usepackage{enumerate}
\usepackage{float} % place figure where you like
\usepackage{textcomp}
\usepackage{upgreek}
\usepackage{rotating}
\usepackage{multirow}
\usepackage{xspace}
\usepackage{color}
\usepackage{mathtools}
\usepackage{braket}
\usepackage{fancybox,calc}
\usepackage{bm}
\usepackage{cancel}
\usepackage{sty/authblk}% for multiple authors
\usepackage{ctable}
\usepackage{placeins}
\usepackage{mciteplus}
\usepackage[hidelinks]{hyperref}
\usepackage{appendix}
\usepackage{color}
\usepackage{colortbl}
\newboolean{pdflatex}
\setboolean{pdflatex}{false} % use this if using eps figures
\mathtoolsset{showonlyrefs=true}
\mathtoolsset{showmanualtags=true}

\newcommand{\fr}{\dfrac}

\newcommand{\dsps}{\displaystyle}
\newcommand{\drac}[2]{\dsps \frac{\mathrm{d} #1}{\mathrm{d} #2}}
\newcommand{\prac}[2]{\dsps \frac{\partial #1}{\partial #2}}
\newcommand{\pr}{\prac}
\newcommand{\executeiffilenewer}[3]{%
\ifnum\pdfstrcmp{\pdffilemoddate{#1}}%
{\pdffilemoddate{#2}}>0{\immediate\write18{#3}}
\fi}

\IfFileExists{InkscapeCommand.tex}%
{}
{}

\graphicspath{{figures/}}
\definecolor{lightgray}{gray}{0.95}

\begin{document}

\title{Signal Formation in a Detector with one Large Dimension}
\renewcommand\Authands{, } % avoid ``. and'' for last author
\renewcommand\Affilfont{\itshape\small} % affiliation formatting

\author[]{Manolis Dris}

\affil[]{National Technical University of Athens, Department of Physics,\\
9 Heroon Polytechniou Street, GR 157 80, Athens, Greece}

%%%%%%%%%%%%%%%%%%%%%%%%%%%%%%%%%%%%%%%%%%%%%%%%%%%%%%%%%%%%%%%%%%%%%%%%%%%%%%%%

\maketitle

\begin{abstract}
\noindent We present the theory for the signal formation in a multi
conductor detector with cylindrical geometry and long length. There
exists electromagnetic wave propagation along the large dimension of
the detector. The system is equivalent to a multi conductor
transmission line. The treatment is in the TEM approximation. Each
conductor is fed by its current source which is the same as in the
case of small size detectors. A simple example is given for a long
length Monitored Drift Tube (MDT). One could apply the result to a
long micromegas-type detector or any long microstrip detector,
ignoring propagation that is transverse to the strips.
\end{abstract}

\tableofcontents

%==========================Introduction==========================
%%%%%%%\input{Introduction}
%================================================================

\input{general_theory}

%==========================Introduction==========================
\input{Conclusions}

%================================================================

%==========================Acknowledgements=======================
\input{Acknowledgments}
%==================================================================

%==========================Bibliography========================

\input{bibliography}
%==========================================================

\end{document}

%% file: general_theory.tex
\section{Signal formation in a detector with cylindrical geometry and long length}\label{sec:tempropdet}

\noindent The problem of induced currents on conducting electrodes
due to the motion of electrons in between the electrodes' vacuum
space, dates back to the 1930's and 1940's . At that time various
types of vacuum tube devices were in use and such effects were
important at high enough frequencies, when the electron time of
flight between the electrodes was comparable to the period of the
radiofrequencies involved (see the classic papers by W. Shockley
~\cite{shockley} and by S. Ramo ~\cite{ramo}). Following similar
 techniques, the problem of signal formation in particle  detectors is
 analysed in several papers and books,
  ~\cite{rossi, gatti, radeka, blum, dris}.
In all cases, small size detectors is ussumed, since electrostatics
is used with no electromagnetic wave propagation. There are
applications of the above techniques for the case of long length
detectors where wave propagation exists along the detector length,
as in ~\cite{schneider}. As far as we know, no rigorous
justification exists for doing so. In this work we give a rigorous
proof of what happens for loong length detectors.

\begin{figure} [htb]
  \centering
  \input{#figures/transvsys.pdf_tex}
  \caption{Cross-section of a system of $N$ internal conductors and one external
  conductor, forming a long-length detector. The geometry is cylindrical.}\label{fig:mmsr-2-1}
\end{figure}
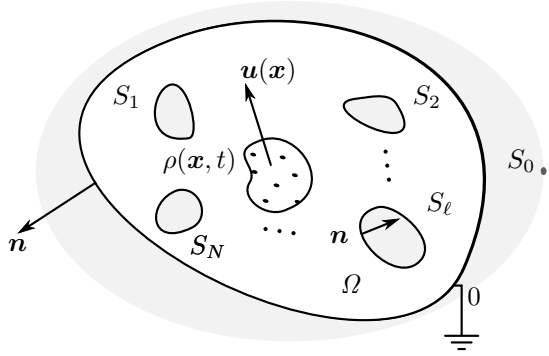

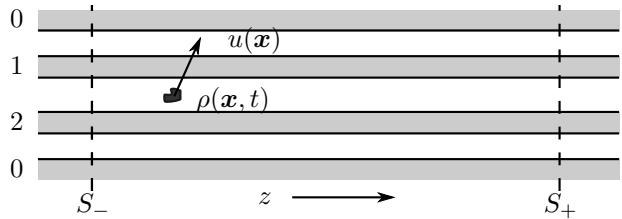
\begin{figure}[!h]
  \centering
  \input{#figures/longitudinal.pdf_tex}
  \caption{Longitudinal view of a transmission line with many conductors. Only
  two internal conductors are shown and the external conductor of zero potential. The
  shaded part shows the inside of the conductors, whereas the part in white is
  the space where the waves propagate. There is a time-dependent (moving
  charge) in between conductors.}\label{fig:mmsr-2-2}
\end{figure}

\noindent The cylindrical geometry of the detector is shown in Fig.
 \ref{fig:mmsr-2-1} and Fig. \ref{fig:mmsr-2-2}. We will start by examining an ideal detector which consists of
many parallel conductors without resistance. The criterion for a
material to be a very good conductor, is the relaxation time $ \tau
= \epsilon / \sigma $ (i.e. permittivity divided by conductivity) to
be much smaller than the periods ($ T=1/f$) of the waves involved.
If the opposite is true, then the material behaves more like a
dielectric. Between the two extremes one has dielectric materials
with conductivity.

\noindent First we assume the space between the conductors contains
a homogeneous linear dielectric medium whose permittivity does not
depend on frequency, i.e. $\epsilon = \epsilon_\textrm{r} \epsilon
_0=\text{constant}$. The medium could be a gas. The motion of a
charge in the space between the ideal conductors, excites the system
and as a result signals are formed and propagate to the ends of the
conductors, where they are detected by the external circuits
connected. We examine an ideal case without any dielectric "losses".
This means there is no any conduction (transverse) current in the
dielectric and there are no dielectric polarization losses.

\noindent The transverse dimensions of the detector are small, such
that there is no electromagnetic wave propagation in the transverse
plane, instead, propagation occurs along the length of the
cylindrical detector (the $z$-axis). It is assumed that the
transverse dimensions are small in comparison to the wavelengths
involved in the problem. If one wants to examine even a small width
plane detector with long parallel conductors, extending transversely
in a large distance, our analysis is not applicable. In that case,
there is propagation in both directions of the plane detector.

\noindent It is clear that we have to solve the problem of
excitation of a  multi conductor  transmission line by moving
charges in between the  conductors' space. We will see that it is
enough to assume that the conditions for electromagnetic wave
propagation of type TEM (Transverse Electromagnetic Waves) apply. In
this case, the fields $\bm{E}$, $\bm{B}$, $\bm{D}$ and $\bm{H}$ are
transverse. The TEM waves are also called type-T (Transverse) or EM
waves. The transmission line system consists of two or more total
number of conductors 0, 1, 2,...$N$. In the study of excitation of
transmission lines (the same is true for wave guides and
electromagnetic cavities, but in those last two cases no TEM modes
exist), we must take into account all modes of excitation, including
the evanescent modes which are not propagating modes. This is
necessary when there is interest to calculate the impedance the
excitation source "sees". In this case, the stored energy and the
energy that returns to the excitation source becomes significant. In
practice, the read out of the detector conductor is "far away" from
the area of the excitation (position of passage of the particle),
meaning at distances of the order of several detector gaps. Note
that the evanescent mode strengths decay quickly with distance, so
they can be ignored. At the same time the modes with evanescence, TE
(or M) and TM (or E), for the usual detectors, have very high
thresholds for their cutoff frequencies, with propagation occuring
only above these cutoffs. These cutoffs are of the order of tenths
of GHz and more. Modeling usual detector signals is accurate enough
with the use of much lower frequencies. The TEM modes can describe
the detector signals very well. They do not have cutoff frequencies,
so in principle, even the very low frequencies are contributing to
this type of transmission. The idea of using only TEM propagation is
supported from the fact mentioned before, that the wavelengths of
the involved waves are much larger than the various transverse
dimensions. One additional reason for using only the TEM modes, is
that the electronics used are working at much lower frequencies than
the above cutoff frequencies. It is worth mentioning that only for
the case of the TEM transmission there is a clear meaning to the
potential difference between two conductors. There is no such clear
meaning for the TE and TM cases.

\noindent There exist detectors having, several (instead of one)
homogeneous dielectric materials with different permittivities,
distributed such that still cylindrical geometry detector is still
homogeneous along its length. These are detectors with inhomogeneous
media. In addition, the conductors may have significant resistances
which influences the system in various ways. In this case our
analysis fails. The fields are not exactly transverse, and have
longitudinal components along the $z$ axis. If permittivities are
not extremely different, and resistances are not very high, one
could apply the quasi-TEM method, where the fields are almost of the
TEM type, with small longitudinal components. The quasi-TEM method
leads to similar relations like the TEM case, with the difference
that several propagation "constants" (propagation coefficients) and
so propagation speeds may appear. Sometimes the propagation
coefficients and speeds do not differ much from each other, so one
could treat the problem as one with a single propagation coefficient
and one speed. One could take an average of permittivities and use
only one value. The maximum number of propagation coefficients is in
general equal to the number of (internal) conductors $(N)$,
~\cite{marx}.

\noindent At this point it should be noted that, even when there are
currents in the dielectric, there are exact TEM waves. This is true
if the dielectric is homogeneous and the conductors are ideal.

\noindent Rossi ~\cite{rossi} seems to be the first who examined the
behavior of a long length detector. Such detectors are in use in
High Energy Physics experiments, e.g. in the ATLAS experiment at
CERN ~\cite{atlas,alexopoulos}.

\noindent We proceed with the lossless transmission line equivalent
of the detector, i.e. conductors with no resistance and dielectric
without any "lossess", i.e.  no conductivity and polarization
losses. The conductors, $N$ internal and the surrounding 0th
conductor at zero potential, could be extend from $z<0$ to $z>0$.
One or both extensions is assumed to be sufficiently large. The
surrounding conductor may partially or totally include infinity. We
use cartesian coordinates for positions $\bm{x}
=(\bm{x}_{\textrm{T}} , z)= (x, y, z)$. Vectors
$\bm{x}_{\textrm{T}}= (x, y)$ are in the transverse plane, normal to
the $z$-axis. The infinitesimal area on the transverse plane is
$\mathrm{d}^2x= \mathrm{d}a= \mathrm{d}x \mathrm{d}y$. The
infinitesimal volume is $\mathrm{d}^3x= \mathrm{d}x \mathrm{d}y
\mathrm{d}z$.

\noindent The time-dependent (moving) charge distribution, in the
space between the conductors, will excite the system. In what
follows we mainly use Jackson, ~\cite{jackson}. The other references
used are ~\cite{collin,borgnis}. The following relations
%\eqref{eq:mmsr-2-1}% hold

\begin{equation}
  \rho = \rho (\bm{x}, t), \quad \bm{J} = \rho (\bm{x}, t) \bm{u} (\bm{x})
  \label{eq:mmsr-2-1}
\end{equation}

\noindent We assume that the velocity of the distributed charge is
known at each time for every point in space. In general the
excitation leads to signals propagating towards both, opposite to
each other, $z$ directions.

\noindent Maxwell's equations in space with a dielectric without
losses ($\bm{J}=0, \ \rho =0$), can be written as follows

\begin{gather}
  \begin{aligned}
    \bm{\nabla}\times \bm{H}&= \epsilon \pr{\bm{E}}{t}, \quad
    \bm{\nabla}\times \bm{E}= -\mu  \pr{\bm{H}}{t}\\
    \bm{\nabla}\cdot \bm{E}&=0, \quad \bm{\nabla}\cdot \bm{H}=0, \quad \bm{D}=
    \epsilon  \bm{E}, \quad \bm{B}= \mu \bm{H}
    \label{eq:mmsr-2-2}
  \end{aligned}
\end{gather}

\noindent It is understood that the appropriate boundary conditions
exist, on the ideal conductors.

\noindent Remember that for TEM waves, the fields are transverse to
the $z$ propagating direction, in the between the conductors space.

\noindent We Fourier transform the various physical quantities and
end up with the quantities called amplitudes, such as $\bm{E}
(\bm{x}, \omega )$. This way we reduce the problem to the one
frequency case. The following relations hold

\begin{gather}
  \begin{aligned} \label{eq:mmsr-2-3}
  \bm{E} (\bm{x}, t) &= \fr{1}{2\pi } \int_{-\infty}^{\infty} \mathrm{d}\omega
  \bm{E} (\bm{x}, \omega ) e^{-j\omega t}, \qquad \bm{E} (\bm{x},
  \omega) = \int_{-\infty}^{+\infty}  \bm{E} (\bm{x}, t)
  e^{j\omega t} \mathrm{d}t \\
  \bm{H} (\bm{x}, t) &= \fr{1}{2\pi } \int_{-\infty}^{\infty}
  \bm{H} (\bm{x}, \omega) e^{-j\omega t}\mathrm{d}\omega, \qquad \bm{H} (\bm{x},
  \omega)= \int_{-\infty}^{+\infty} \bm{H} (\bm{x}, t)
  e^{j\omega t}\mathrm{d}t  \\
  \rho (\bm{x}, t) &= \fr{1}{2\pi } \int_{-\infty}^{\infty}
  \mathrm{d}\omega \rho  (\bm{x}, \omega) e^{-j\omega t}, \qquad \rho
  (\bm{x}, \omega)= \int_{-\infty}^{+\infty} \rho  (\bm{x}, t)
  e^{j\omega t} \mathrm{d}t \\
  \bm{J} (\bm{x}, t) &= \fr{1}{2\pi } \int_{-\infty}^{\infty}
  \bm{J} (\bm{x}, \omega) e^{-j\omega t}\mathrm{d}\omega, \qquad \bm{J} (\bm{x},
  \omega)= \int_{-\infty}^{+\infty}\bm{J} (\bm{x},
  t)e^{j\omega t}\mathrm{d}t
\end{aligned}
\end{gather}

\noindent The continuity relations are %Eqs \eqref{eq:mmsr-2-4}%

\begin{equation}
  \bm{\nabla} \cdot \bm{j} (\bm{x}, t) + \pr{\rho  (\bm{x}, t) }{t}=0, \qquad
  \bm{\nabla}\cdot \bm{J} (\bm{x}, \omega)= j\omega \rho  (\bm{x}, \omega)
  \label{eq:mmsr-2-4}
\end{equation}

\noindent Substitution into Maxwell Eqs. \eqref{eq:mmsr-2-2} yields
%Eqs \eqref{eq:mmsr-2-5}%

\begin{gather}
  \begin{aligned}
  \label{eq:mmsr-2-5}
   \bm{\nabla}\times  \bm{H} (\bm{x}, \omega)+ j\epsilon \omega \bm{E}
   (\bm{x}, \omega)&= 0 \\
   \bm{\nabla}\times \bm{E}(\bm{x}, \omega)- j\mu \omega \bm{H} (\bm{x},
   \omega)&=0  \\
   \bm{\nabla}\cdot \bm{H} (\bm{x}, \omega)=0, \qquad \bm{\nabla}\cdot \bm{E}
   (\bm{x}, \omega)&= 0
  \end{aligned}
\end{gather}

\noindent From these we get the wave equations %\eqref{eq:mmsr-2-6}%

\begin{gather} \label{eq:mmsr-2-6}
  \begin{aligned}
    \nabla^2 \bm{E} (\bm{x}, \omega)+\mu \epsilon \omega ^2 \bm{E} (\bm{x},
    \omega)&=0 \\
    \nabla^2 \bm{H} (\bm{x}, \omega)+\mu \epsilon \omega ^2 \bm{H} (\bm{x},
    \omega)&=0
  \end{aligned}
\end{gather}

\noindent The geometry is cylindrical with propagation only along
the $z$-axis, so we assume solutions of the form %as in Eqs
\eqref{eq:mmsr-2-7}%

\begin{gather}
  \begin{aligned}
  \label{eq:mmsr-2-7}
    \bm{E} (\bm{x}, \omega)&= \bm{E}_0(\bm{x}_{\textrm{T}},
    \omega)e^{(\pm) jkz}\\
    \bm{H} (\bm{x}, \omega)&= \bm{H}_0(\bm{x}_{\textrm{T}},
    \omega)e^{(\pm) jkz} \\
    \bm{x}_{\textrm{T}}&= (x, y)
  \end{aligned}
\end{gather}

\noindent The plus sign in the exponential, corresponds to a
propagation along the positive $z$-direction, and the minus sign
corresponds to the negative $z$-direction. We get relations

\begin{gather}
\begin{aligned}
\label{eq:mmsr-2-8}
  E_z&=0, \qquad H_z=0 \\
  \bm{\nabla}_{\textrm{T}}\times \bm{E}_0&=0, \qquad
  \bm{\nabla}_{\textrm{T}}\cdot \bm{E}_0=0, \qquad
  \bm{\nabla}_{\textrm{T}}= \bm{e}_x \pr{}{x}+ \bm{e}_y \pr{}{y} \\
  \bm{H}_0&= (\pm) \fr{1}{Z_{\textrm{T}}} \bm{e}_z\times  \bm{E}_0, \qquad
  Z_{\textrm{T}}= \sqrt{\dfrac{\mu }{\epsilon }}= \text{the wave impedance}
   \\
  k&=\omega \sqrt{\mu \epsilon }= \fr{\omega }{c}\approx \omega
  \sqrt{\mu _0\epsilon _0\epsilon _{\textrm{T}}}, \quad
  c=1/\sqrt{\mu \epsilon }= \text{speed of light in the medium}
\end{aligned}
\end{gather}

\noindent Wavenumber $k$ coincides with that of the unconfined
medium. The speed of light in vacuum is $c_0=1/\sqrt{\mu _0\epsilon
_0}$, so $c=c_0/\sqrt{\mu _{\textrm{r}} \epsilon _{\textrm{r}}
}\approx c_0/\sqrt{\mu _0\epsilon _{\textrm{r}} }$.

\noindent Whenever the +,- signs exist in an expression more than
once, the upper signs go together and the down signs go together.
Eqs. \eqref{eq:mmsr-2-8} expresses the known fact for cylindrical
geometry,  that we have electrostatic relations in two dimensions,
in the transverse plane $(x, y)=\bm{x}_{\textrm{T}}$. This means
that electric field $\bm{E}_0(\bm{x}_{\textrm{T}})$, which is
transverse, can be determined from the two-dimensional scalar
electrostatic potential ${\it\Phi} _0(\bm{x}_{\textrm{T}})$, as
shown below in Eqs. \eqref{eq:mmsr-2-9}, with the appropriate
boundary conditions.

\begin{gather}\label{eq:mmsr-2-9}
  \begin{aligned}
    \bm{E}_0 (x, y)&= - \bm{\nabla}_{\textrm{T}} {\it\Phi} _0 (x, y) \\
    \nabla^2_{\textrm{T}}{\it\Phi} _0 (x, y)&= 0
  \end{aligned}
\end{gather}

\noindent Since the conductors are ideal (with no resistance), the
electric field, in the outside the conductors space, is normal to
the conductor surfaces and the magnetic field intensity outside each
conductor and very close to it, is parallel to the conductor
surface. It should be noted that, in practice, the materials used
are non-magnetic, so $\mu _{\textrm{r}} \approx 1$.

\noindent Maxwell's equations with the appropriate boundary
conditions, for the transmission lines, define a problem of
eigenvalues. The eigensolutions of the problem are the normal modes
of transmission, and constitute a complete set of solutions. All the
modes (TE, TM, TEM) exist in the case of the transmission lines. So
in the case of our detectors all modes contribute. Any
electromagnetic field inside the detector, of one frequency, with
the correct boundary conditions, can be represented as a sum of the
above complete set of solutions with appropriate coefficients. This
is the method used for the description of the excitation of wave
guides and electromagnetic cavities where no TEM modes exist,
~\cite{jackson,collin,borgnis}. We will make use of this method, for
the excitation of transmission lines from the motion of a charge
distribution, adapting it for our case where the TEM mode exists and
it is the only one.This means we will use only the eigensolutions
TEM, not the complete set. We have to solve the two dimensional
Laplace equation for the potential, with the appropriate boundary
conditions. The equation is the second from relations
\eqref{eq:mmsr-2-9}. The solution is ${\it\Phi}_0(x,y)$ which on the
surface of each one conductor, $\lambda$, will have potential
$V_\lambda$. Let us consider a cylindrical three dimensional space
with cross section as in Fig.1 and small height $\Delta z$. This
space is part of the cylindrical space of large height. We will
consider that symbols $\it{\Omega},S_0,S_1,S_2,...,S_N$ in Fig.
\ref{fig:mmsr-2-1} are referring to this small cylindrical space.
The small cylindrical space $\it\Omega$ is enclosed by part of the
cylindrical surfaces of all the conductors, and two imaginary plane
surfaces normal to the $z$ axis, with the between them distance
(height of the small cylinder) $\Delta z$. As we said before for TEM
mode transmission, the problem for the transverse directions is
reduced to an electrostatic problem. We will see later on that
application of the Gauss theorem etc, one concludes that between the
(electrostatic) potentials $V_\lambda$ for conductors
$\lambda=1,2,...,N$ and the corresponding charges $Q_\lambda$, the
following relations hold

\begin{gather} \label{eq:mmsr-2-10}
\begin{aligned}
  Q_j&= \sum_{\lambda =1}^{N} Q_{j\lambda } = \sum_{\lambda =1}^{N}c_{j\lambda } V_\lambda , \quad j=1, \ldots, N \\
  c_{j\lambda }&= c_{j\lambda }\leq 0\quad \forall j\neq \lambda , \quad c_{jj}\geq
  0 \\ U&= \fr12 \sum_{i,j=1}^{N} c_{ij} V_iV_j
\end{aligned}
\end{gather}

\noindent  $c_{j\lambda }$ is defined by  $c_{j\lambda }=Q_{j\lambda
 } / V_{\lambda }$. $Q_{j\lambda }$ is the induced charge on
conductor $j$ when only conductor $\lambda $  has non zero potential
equal to $V_{\lambda }$ while all other conductors have zero
potential. $U$ is the electrostatic energy of the system.

\noindent From these relations, dividing by the height $\Delta x$,
we find the per unit length corresponding quantities. We will not
change symbols, we will consider that the appropriate quantities in
Eqs. \eqref{eq:mmsr-2-10} are per unit length. We remind that for
this calculation there is not charge in the space in between the
conductors. $c_{jl}$ are the coefficients of electrostatic induction
per unit length, another name is short circuit capacitances per unit
length. $V_\lambda$ are the electrostatic potentials of the various
conductors with respect to the reference conductor (zero potential)
$S_0$. Coefficients $c_{jl}$ are related to the more usual
coefficients $C_{jl}$ appearing in electrical circuits, which are
called capacitances with two terminals, per unit length, or simply
capacitances per unit length. Capacitances with two terminals
constitute the respective matrix $[C]$. These last coefficients are
related to the potential differences between the nodes and for this
reason lead to equivalent electric circuits with capacitances
connected between the various nodes, making the problem easy to
solve with the known methods for electrical circuits. All above
capacitances are the ones used in the description of multi conductor
transmission lines, see the various references from ~\cite{bewley}
to ~\cite{paul}. We have %Eqs \eqref{eq:mmsr-2-11}%

\begin{gather}\label{eq:mmsr-2-11}
\begin{aligned}
  Q_j &= C_{jj} V_j +\sum_{l=1}^{N} C_{jl} (V_{j}-V_{l}), \qquad j=1, \ldots, N
  \\
  C_{jl}&= C_{lj}=- c_{lj}=-c_{jl} \geq 0\quad \forall j\neq l, \qquad C_{ll}=
  \sum_{j=1}^{N} c_{ij} \geq 0  \\
  c_{ii}&= \sum_{k=1}^{N}  C_{ik}, \qquad c_{ij}= -C_{ij}\quad \forall i\neq j
\end{aligned}
\end{gather}

We can invert the first of Eqs. \eqref{eq:mmsr-2-10} and end up with
%Eqs. \eqref{eq:mmsr-2-12}%

\begin{gather}
\begin{aligned}
  V_{j}&= \sum_{l=1}^{N} p_{jl}Q_{l}, \qquad j=1, \ldots, N \\
  p_{jl}&= p_{lj}\geq 0, \qquad p_{jj}\geq p_{il}  \\
  U&= \fr12 \sum_{i,j=1}^{N} p_{ij}Q_{i}Q_{j}
  \label{eq:mmsr-2-12}
\end{aligned}
\end{gather}

\noindent  $U$ is the electrostatic energy per unit length of the
system.

\noindent Coefficients $p_{ij}$ are called (Maxwell) potential
coefficients per unit length. Let $[c]$ be the (NxN) matrix of
coefficients $c_{ij}$ and $[P]$ be the (NxN) matrix of $p_{ij}$, we
then have %Eqs \eqref{eq:mmsr-2-13}%

\begin{equation}\label{eq:mmsr-2-13}
  [c]=[p]^{-1}, \qquad [p]= [c]^{-1}, \qquad [c][p]= [E_{\textrm{u}}]
\end{equation}

\noindent If we have $N$ (internal) conductors, as shown in Fig.
\ref{fig:mmsr-2-1}, to solve the problem of signal formation, we
will  proceed as follows.

\noindent We imagine that only one conductor, $\lambda $, has
potential $V_\lambda \neq 0$, while all the rest have zero
potential. We solve the $N$ electrostatic problems and determine the
respective potentials ${\it\Phi}_{0\lambda }(x, y)$. It is clear
that in the case the conductors, simultaneously  have potentials
$V_{\lambda }$, $\lambda =1, 2, \ldots, N$, respectively, we will
have ${\it\Phi} _0(x, y)= \sum_{\lambda =1}^{N} {\it\Phi} _{0\lambda
}(x, y)$.

\noindent For the electric eigenfield $\bm{E}_{0\lambda }(x, y)$ we
have %the following Eqs \eqref{eq:mmsr-2-14}%

\begin{gather} \label{eq:mmsr-2-14}
  \begin{aligned}
    \bm{E}_{0\lambda } (x, y) &= -\bm{\nabla}{\it\Phi} _{0\lambda }(x, y) \\
    \bm{E}_0 (x, y) &= -\bm{\nabla}{\it\Phi} _0(x, y) = \sum_{\lambda =1}^{N}
    \bm{E}_{0\lambda } (x, y)
  \end{aligned}
\end{gather}

\noindent The eigenfields $\bm{E}_{0\lambda } (x, y) $ are taken to
be real. For their values  in every point in the between the
conductors space, and for one frequency, we have the relations:

\begin{gather}
  \begin{aligned}
    \bm{E}_{0\lambda }(\bm{x}_\textrm{T})&= e^{(\pm) jkz}
    \bm{E}_{0\lambda } (\bm{x}_{\textrm{T}})\\
    \bm{H}^{\pm}_{0\lambda } (\bm{x}_{\textrm{T}})&= \mp
    e^{(\pm)jkz} \\
    \bm{H}_{0\lambda }(\bm{x}_{\textrm{T}})&= \mp \fr{1}{Z_\textrm{T} }
    \bm{e}_z\times \bm{E}_{0\lambda }(\bm{x}_{\textrm{T}}), \quad k= \fr{\omega
    }{c}
    \label{eq:mmsr-2-15}
  \end{aligned}
\end{gather}

\noindent The "intensities" of the electric and magnetic field for
any propagating electromagnetic wave, will depend on the above
eigenfields. We have

\begin{gather}
  \begin{aligned}
    \bm{E}^{(\pm)} (\bm{x}, \omega)&= \sum_{\lambda =1}^{N} A^{(\pm)}_\lambda
    (\omega ) \bm{E}^{(\pm)}_{0\lambda } (\bm{x}_{\textrm{T}}, z)= \sum_{\lambda
    =1}^{N} A^{(\pm)}_\lambda  (\omega ) \bm{E}_{0\lambda }
    (\bm{x}_{\textrm{T}}) e^{(\pm)jkz} \\
    \bm{H}^{(\pm)} (\bm{x}, \omega)&= \sum_{\lambda =1}^{N} A^{(\pm)}_\lambda
    (\omega ) \bm{H}^{(\pm)}_{0\lambda } (\bm{x}_{\textrm{T}}, z)= \mp \sum_{\lambda
    =1}^{N} A^{(\pm)}_\lambda  (\omega ) \bm{H}_{0\lambda }
    (\bm{x}_{\textrm{T}}) e^{(\pm)jkz} \\
    &= \mp \fr{1}{Z_\textrm{T} }\bm{e}_z\times \bm{E}^{(\pm)} (\bm{x}, \omega),
    k= \fr{\omega }{c}\\
    \textrm{In general,}\\
  \bm{E} (\bm{x}, \omega) &= \bm{E}^{(+)} (\bm{x}, \omega)+ \bm{E}^{(-)} (\bm{x},
  \omega)\quad \textrm{and} \quad \bm{H}(\bm{x}, \omega)= \bm{H}^{(+)} (\bm{x},
  \omega)+ \bm{H}^{(-)} (\bm{x}, \omega)
   \label{eq:mmsr-2-16}
  \end{aligned}
\end{gather}

\noindent Finding coefficients $A^{(\pm)}_{\lambda }(\omega )$
determines the fields at any point in the space we are interested
in.

\noindent It can be proven that, if the fields are known in any
plane normal to the $z$-axis, this determines completely the two
coefficients, so the fields can be found at anyone point. We select
as a transverse plane, the one at $z=0$.

\noindent Let us assume that the transmission line is excited by a
localized source, as shown in Figs \ref{fig:mmsr-2-1} and
\ref{fig:mmsr-2-2}.

\noindent We use the divergence theorem and we get the following
relations for the potentials and the eigenfields

\begin{gather} \label{eq:mmsr-2-17}
\begin{aligned}
  \bm{\nabla}\cdot\left({\it\Phi} _{0\lambda } \bm{\nabla} {\it\Phi}
  _{0\lambda^{\prime}}\right)&= \bm{\nabla} {\it\Phi}_{0\lambda}\cdot
  {\it\Phi}_{0\lambda^{\prime}}+{\it\Phi}_{0\lambda}\cdot\nabla^2
  {\it\Phi}_{0\lambda^{\prime}}, \quad \nabla^2 {\it\Phi}_{0\lambda^{\prime}}=0 \\
  \int_{\it\Omega}  \bm{E}_{0\lambda}\cdot \bm{E}_{0\lambda^{\prime}} \mathrm{d}^3x&=
  \int_{\it\Omega}  \bm{\nabla} {\it\Phi}_{0\lambda}\cdot
  {\it\Phi}_{0\lambda^{\prime}}\mathrm{d}^3x= \int_{\it\Omega}  \bm{\nabla}\cdot\left(
  {\it\Phi}_{0\lambda}\bm{\nabla}{\it\Phi}_{0\lambda^{\prime}} \right)\mathrm{d}^3x \\
  \Delta z \int_S \bm{E}_{0\lambda}\cdot\bm{E}_{0\lambda^{\prime}} \mathrm{d}a&=
  \sum_{m=1}^{N} \int_{S_m}
  {\it\Phi}_{0\lambda}\bm{\nabla}{\it\Phi}_{0\lambda^{\prime}}\bm{e}_z \mathrm{d}a=
  V_\lambda \int_{S_\lambda }\bm{\nabla}{\it\Phi}_{0\lambda^{\prime}}\cdot \bm{e}_z
  \mathrm{d}a \\
  &= -V_\lambda  \int_{S_\lambda } \fr{\sigma_{\lambda
  \lambda^{\prime}}}{\epsilon }\mathrm{d}a= - \fr{1}{\epsilon } V_\lambda
  Q_{\lambda \lambda ^{\prime}}= - \fr{1}{\epsilon } V_\lambda  c_{\lambda
  \lambda ^{\prime}} V_{\lambda ^{\prime}}\\
  \int_S \bm{E}_{0\lambda}\cdot\bm{E}_{0\lambda^{\prime}}\mathrm{d}a&=
  \fr{1}{\epsilon } \fr{c_{\lambda \lambda ^{\prime}}}{\Delta z} V_\lambda
  V_{\lambda ^{\prime}}, \qquad \lambda , \, \lambda ^{\prime}= 1, 2, \ldots,
  N \\
  \textrm{hence,} \qquad \qquad \qquad \\
  \int_S \bm{H}_{0\lambda}\cdot \bm{H}_{0\lambda^{\prime}}\mathrm{d}a&=
  \fr{1}{Z^2_\textrm{T} \epsilon } \fr{c_{\lambda \lambda ^{\prime}}}{\Delta
  z} V_\lambda V_{\lambda ^{\prime}} \\
  \int_S \left( \bm{E}_{0\lambda}\times \bm{H}_{0\lambda^{\prime}} \right)\cdot
  \bm{e}_z \mathrm{d}a&= \fr{1}{Z_\textrm{T} \epsilon } \fr{c_{\lambda
  \lambda ^{\prime}}}{\Delta z} V_\lambda  V_{\lambda ^{\prime}}
  \end{aligned}
\end{gather}

\noindent Remember, we use symbol $c_{ij}$ for the capacity per unit
length coefficients, i.e. $c_{ij}/\Delta z\to c_{ij}$. We have
$c_{\lambda \lambda^{\prime}}=Q_{\lambda
\lambda^{\prime}}/V_{\lambda^{\prime}}$. $Q_{\lambda
\lambda^{\prime}}$ is the charge induced on conductor $\lambda $
when only conductor $\lambda^{\prime}$ has a non zero potential
$V_{\lambda^{\prime}}$ while all other conductors have  zero
potentials.

\noindent In Fig. \ref{fig:mmsr-2-2}, outside the source, on the
right of the plane surface $S_+$ that is normal to the $z$-axis in
position $z=z_+$, fields $\bm{E}, \ \bm{H}$ are given, following
Eqs. \eqref{eq:mmsr-2-16}, from %Eqs \eqref{eq:mmsr-2-18}%

\begin{gather}\label{eq:mmsr-2-18}
  \begin{aligned}
    \bm{E}&= \bm{E}^{(+)} (\bm{x}, \omega)&= \sum_{\lambda =1}^{N}
    A^{(+)}_\lambda  (\omega ) \bm{E}_{0\lambda} (\bm{x}_{\textrm{T}})
    e^{(+)jkz} \\
    \bm{H}&= \bm{H}^{(+)} (\bm{x}, \omega)&= \sum_{\lambda =1}^{N}
    A^{(+)}_\lambda  (\omega ) \bm{H}_{0\lambda} (\bm{x}_{\textrm{T}})
    e^{(+)jkz}
  \end{aligned}
\end{gather}

\noindent On the left of the respective surface $S_-$ ($z=z_-$),
fields $\bm{E}$  are given, again following Eqs.
\eqref{eq:mmsr-2-16}, from

\begin{gather}\label{eq:mmsr-2-19}
  \begin{aligned}
    \bm{E}&= \bm{E}^{(-)} (\bm{x}, \omega)&= \sum_{\lambda =1}^{N}
    A^{(-)}_\lambda  (\omega ) \bm{E}_{0\lambda} (\bm{x}_{\textrm{T}})
    e^{(-)jkz} \\
    \bm{H}&= \bm{H}^{(-)} (\bm{x}, \omega)&= -\sum_{\lambda =1}^{N}
    A^{(-)}_\lambda  (\omega ) \bm{H}_{0\lambda} (\bm{x}_{\textrm{T}})
    e^{(-)jkz}
  \end{aligned}
\end{gather}

\noindent As mentioned above, we follow a similar procedure as in
references, ~\cite{jackson,collin,borgnis}. We make the needed
modifications, because in our case we have only TEM propagation. We
take care of the fact that the eigenfunctions are not orthogonal,
see Eq. \eqref{eq:mmsr-2-17}. We use the Lorentz reciprocity
theorem, ~\cite{collin}, which we write for the ``real'' state with
quantities $\bm{E} (\bm{x}, \omega)$, $\bm{H} (\bm{x}, \omega)$,
$\bm{J} (\bm{x}, \omega)$ and for each one of the eigenstates
$\bm{E}_\lambda ^{(\pm)} (\bm{x}, \omega)$, $\bm{H}^{(\pm)}_\lambda
(\bm{x}, \omega)$, $\bm{J}(\bm{x}, \omega)=0$, $\lambda =1, 2,
\ldots, N$. We get %Eqs \eqref{eq:mmsr-2-20}%

\begin{equation}
  \int_S \left( \bm{E}\times \bm{H}^{(\pm)}_\lambda - \bm{E}^{(\pm)}_\lambda
  \times  \bm{H} \right)\cdot \bm{n}\mathrm{d}a= \int_{\it\Omega}  \left( \bm{J}\cdot
  \bm{E}_\lambda ^{(\pm)} \right)\mathrm{d}^3x, \quad \lambda =1, 2, \ldots, N
  \label{eq:mmsr-2-20}
\end{equation}

\noindent Fields $\bm{E} (\bm{x}, \omega)$, $\bm{H} (\bm{x},
\omega)$ come from the excitation of current $\bm{J} (\bm{x},
\omega)$.  $\bm{E}_\lambda ^{(\pm)} (\bm{x}, \omega)$,
$\bm{H}_\lambda ^{(\pm)} (\bm{x}, \omega)$ are the TEM-type
eigenfields propagating along the positive and negative
$z-$directions, index $\lambda $ characterizes the propagating field
when only conductor $\lambda $ has a non-zero potential while all
the rest have zero potential. The $\it\Omega$ space is the space
between conductors (multiply connected), surrounded by external
conductor $S_0$ and the two imaginary plane surfaces $S_+, S_-$,
Fig. \ref{fig:mmsr-2-2}; this is the space that contains all the
time the (localized) field sources.

\noindent Since we have ideal conductors, the electric field is
normal to their surfaces, thus the surface integral on the
cylindrical part of surface $S$ is zero, so only the integrals on
the two flat surfaces need to be found. We have

\begin{equation}
  \int_{\it\Omega} \left( \bm{J}\cdot \bm{E}_\lambda ^{(\pm)} \right)\mathrm{d}^3x
  = \int_{S^+} \left( \bm{E}\times \bm{H}_\lambda^{(\pm)} - \bm{E}_\lambda
  ^{(\pm)} \times \bm{H} \right)\cdot \bm{n}\mathrm{d}a + \int_{S_-} \left(
  \bm{E}\times  \bm{H}_\lambda  ^{(\pm)} - \bm{E}_\lambda ^{(\pm)} \times  \bm{H}
  \right)\cdot \bm{n} \mathrm{d}a
  \label{eq:mmsr-2-21}
\end{equation}

\noindent Field $\bm{E}$ is a result of excitation in the
transmission line space between surfaces $S_-$ and $S_+$. This means
that on the side of surface $S_+$, $\bm{E}$ we will propagate along
the positive $z$-direction, while on the side of $S_-$ along the
negative $z$-direction.

\noindent We choose to work with the minus signs in Eqs.
\eqref{eq:mmsr-2-21}. In addition, using Eqs. \eqref{eq:mmsr-2-19},
we find that the last surface integral is zero, i.e.
$\int_{S_-}\ldots =0$. Thus, according to Eqs. \eqref{eq:mmsr-2-15}
we have $\bm{H}_{0\lambda}= \mp (1/Z_\textrm{T} ) \bm{e}_z\times
\bm{E}_{0\lambda}$. The identity $\bm{a}\times  (\bm{b}\times
\bm{c})= (\bm{a}\cdot \bm{c}) \bm{b}- (\bm{a}\cdot \bm{b})\bm{c}$
holds, so we find %that Eqs \eqref{eq:mmsr-2-22}  hold%

\begin{equation}
  \int_{\it\Omega} \left( \bm{J}\cdot \bm{E}_\lambda ^{(-)} \right)\mathrm{d}^3x= -
  \fr{2}{Z_\textrm{T} } \sum_{\lambda^{\prime}=1}^{N} A_{\lambda ^{\prime}}
  \int_{S_+} \left( \bm{E}_{0\lambda}\cdot \bm{E}_{0\lambda^{\prime}}
  \right)\mathrm{d}a, \qquad \lambda =1, 2, \ldots, N
  \label{eq:mmsr-2-22}
\end{equation}

Using Eqs. \eqref{eq:mmsr-2-15} we finally get %Eqs
\eqref{eq:mmsr-2-23}%

\begin{equation}
  \int_{\it\Omega} \left( \bm{J}\cdot \bm{E}_\lambda ^{(-)}/ V_\lambda
  \right)\mathrm{d}^3x = - \fr{2}{Z_\textrm{T} \epsilon } \sum_{\lambda
  ^{\prime}=1}^{N}  A_{\lambda ^{\prime}}^{(+)} V_{\lambda^{\prime}} c_{\lambda
  ^{\prime}\lambda }, \quad \lambda =1, 2, \ldots, N
  \label{eq:mmsr-2-23}
\end{equation}

\noindent Similar results hold for the plus sign in Eq.
\eqref{eq:mmsr-2-17}. At the end, we arrive at

\begin{equation}
  \int_{\it\Omega} \left( \bm{J}\cdot \bm{E}_\lambda^{(\mp)}/ V_\lambda
  \right)\mathrm{d}^3x= - \fr{2}{Z_\textrm{T} } \sum_{\lambda ^{\prime}=1 }^{N}
  A_{\lambda ^{\prime}}^{(\pm)}  V_{\lambda ^{\prime}} c_{\lambda
  ^{\prime}\lambda }, \qquad \lambda =1, 2, \ldots, N
  \label{eq:mmsr-2-24}
\end{equation}

\noindent From Eqs. \eqref{eq:mmsr-2-19} we can find the potential
of conductor $\lambda $ with respect to the reference conductor 0,
by integrating the electric field intensity along an arbitrary path
on the plane with constant $z$, from an arbitrary point of the
surface of conductor $\lambda$ to an arbitrary point on the
reference conductor. The value of the voltage for conductor $\lambda
$ is

\begin{equation}
  v_\lambda ^{(\pm)} (z, \omega )= \int_{\lambda }^0 \bm{E}^{(\pm)}  (\bm{x},
  \omega)\mathrm{d} \bm{r}= e^{(\pm)jkz} \sum_{l^{\prime}=1}^{N}
  A_{\lambda ^{\prime}}^{(\pm)} \int_\lambda ^0 \bm{E}_{0\lambda^{\prime}}
  (\bm{x}_{\textrm{T}}) \cdot \mathrm{d} \bm{r}= e^{(\pm)jkz} A_\lambda
  ^{(\pm)} V_\lambda
  \label{eq:mmsr-2-25}
\end{equation}

\noindent We multiply Eq. \eqref{eq:mmsr-2-24} with $e^{(\pm)jkz}=
e^{(\pm)j\omega z/c}$ and, by using Eq. \eqref{eq:mmsr-2-25} we get

\begin{align}
  e^{(\pm)jkz} \int_{\it\Omega}  \left( \bm{J}\cdot \bm{E}_\lambda
  ^{(\pm)}/V_\lambda  \right)\mathrm{d}^3 x &= - \fr{2}{Z_\textrm{T} \epsilon
  } e^{(\pm)jkz} \sum_{\lambda ^{\prime}=1}^{N} A_{\lambda
  ^{\prime}}^{(\pm)} V_{\lambda ^{\prime}} c_{\lambda ^{\prime}\lambda } \\
  &= - \fr{2}{Z_\textrm{T} \epsilon } \sum_{\lambda ^{\prime}=1}^{N}
  v_{\lambda ^{\prime}}^{(\pm)} (z,\omega ) c_{\lambda ^{\prime}\lambda },
  \quad \lambda =1, 2, \ldots, N \label{eq:mmsr-2-26} \\
  Z_\textrm{T} \epsilon &= \fr{1}{c}
\end{align}

\noindent We proceed  to the final solution of the signal formation
problem. We take the inverse Fourier transform on both members of
Eqs. \eqref{eq:mmsr-2-26}. We end up with relations for the
potentials of the conductors, functions of $z, t$. We use Eq.
\eqref{eq:mmsr-2-3} and we get %Eqs \eqref{eq:mmsr-2-27}%

\begin{gather}\label{eq:mmsr-2-27}
\begin{aligned}
  \fr{1}{2\pi } \int_{-\infty}^{+\infty} &\left[
  e^{-j\omega t} \left( e^{^{(\pm)j\omega z/c}} \int_{\it\Omega}
  \left( \bm{J} (\bm{x}^{\prime}, \omega )\cdot \bm{E}_\lambda ^{(\mp)}
  (\bm{x}^{\prime}, \omega )/V_\lambda  \right)\mathrm{d}^3 x^{\prime}\right)
  \mathrm{d}\omega\right]=\\
  &= -2c \sum_{\lambda ^{\prime}=1}^{N} c_{\lambda ^{\prime}\lambda }
  \fr{1}{2\pi } \int_{-\infty}^{+\infty}  e^{-j\omega
  t} v_{\lambda ^{\prime}} ^{(\pm)}  (z, \omega )\mathrm{d}\omega  \\
  &= -2c \sum_{\lambda ^{\prime}=1}^{N} c_{\lambda ^{\prime}\lambda } v_{\lambda
  ^{\prime}} ^{(\pm)} (z, t), \quad \lambda =1, 2, \ldots, N
\end{aligned}
\end{gather}

\noindent The first member becomes %as in Eqs \eqref{eq:mmsr-2-28}%

\begin{gather}\label{eq:mmsr-2-28}
\begin{aligned}
  \fr{1}{2\pi } \int_{-\infty}^{+\infty}
  &\left[
  e^{-j\omega t}\left(
  e^{(\pm) j\omega z/c} \int_{\it\Omega} e^{(\mp)j\omega
  z^{\prime}/c}\left(
  \bm{J} (\bm{x}^{\prime}, \omega)\cdot
  \bm{E}_{0\lambda}(\bm{x}^{\prime}_{\textrm{T}})/V_\lambda
  \right)\mathrm{d}^3 x^{\prime} \right) \right]\mathrm{d}\omega  \\
  &= \int_{\it\Omega}  \left[
  \left(  \bm{E}_{0\lambda} (\bm{x}^{\prime}_{\textrm{T}})/V_\lambda
  \right)\cdot \left(  \fr{1}{2\pi } \int_{-\infty}^{+\infty}
  \bm{J} \left(  \bm{x}^{\prime}, \omega  \right)e^{-j\omega \left(
  t\mp \fr{z-z^{\prime}}{c} \right)}\mathrm{d}\omega
  \right)\right]\mathrm{d}^3x^{\prime} \\
  &= \int_{\it\Omega}  \left( \bm{E}_{0\lambda}
  (\bm{x}^{\prime}_{\textrm{T}})/V_\lambda  \right)\cdot \bm{J} \left(
  \bm{x}^{\prime}, t \mp \fr{z-z^{\prime}}{c}
  \right)\mathrm{d}^3 x^{\prime}
\end{aligned}
\end{gather}

\noindent Finally we get %Eqs \eqref{eq:mmsr-2-29}%

\begin{equation}
  \fr{-1}{2} \int_{\it\Omega}  \left(
  \bm{E}_{0\lambda}(\bm{x}^{\prime}_\textrm{T} )/V_\lambda  \right)\cdot \bm{J}
  \left(  \bm{x}^{\prime}, t \mp \fr{z-z^{\prime}}{c} \right)\mathrm{d}^3 x^{\prime}= c \sum_{\lambda
  ^{\prime}=1}^{N}  c_{\lambda ^{\prime}\lambda } v_{\lambda ^{\prime}}^{(\pm)}
  (z, t), \quad \lambda =1, 2, \ldots, N
  \label{eq:mmsr-2-29}
\end{equation}

\noindent Since $\bm{E}_\lambda $ is transverse, evidently only the
transverse current density $\bm{J}_\textrm{T} $ contributes to the
signal formation. More precisely, it is the projection of
$\bm{J}_\textrm{T} $ on $\bm{E}_\lambda $'s direction that
contributes, so we get

\begin{equation}
  \fr{-1}{2}\int_{\it\Omega}  \left( \bm{E}_{0\lambda}
  (\bm{x}^{\prime}_{\textrm{T}})/V_\lambda  \right)\cdot \bm{J}_\textrm{T}
  \left(  \bm{x}^{\prime}, t\mp \fr{z-z^{\prime}}{c} \right)\mathrm{d}^3x^{\prime}= c \sum_{\lambda
  ^{\prime}=1}^{N} c_{\lambda ^{\prime}\lambda }v_{\lambda ^{\prime}}^{(\pm)}
  (z, t), \quad \lambda =1, 2, \ldots, N
  \label{eq:mmsr-2-30}
\end{equation}

\noindent If needed, we can translate the origin of time such that
$t \to t- t_{0}$ . This means that excitation starts not at $t=0$
but at $t=t_{0}$ .

\noindent The direction of $\bm{J}_\textrm{T}$ depends on magnetic
field that might be present, so the value of the integral will be
affected by the magnetic field.

\noindent At this  point we are going to study the case of a moving
point charge.

\noindent We assume that we know the travel path of the point charge
as a function of time and its velocity as a function of its
position,  $\bm{x}= \bm{x}_q(t)$,
$\bm{u}=\bm{u}\left(\bm{x}_q\right)= \bm{u}\left( \bm{x}_q \right)=
\left( u_x(\bm{x}_q), u_y(\bm{x}_q), u_z(\bm{x}_q) \right)$.We also
assume that we can express the two position coordinates of the point
charge as a function of one of the two transverse ones, let that be
$x_q$. Index $q$ denotes that the respective quantity refers to the
moving point charge.

\noindent The charge and current densities are given by %Eqs
\eqref{eq:mmsr-2-31}%

\begin{gather}\label{eq:mmsr-2-31}
  \begin{aligned}
    \rho &= \rho (\bm{x}^{\prime}, t) = q\delta\left( \bm{x}^{\prime}-
    \bm{x}_q(t) \right) \\
    \bm{J}&= (J_x, J_y, J_z)= \rho
    \bm{u}= q\delta\left(  \bm{x}^{\prime}- \bm{x}_q(t) \right) \bm{u}\left(
    \bm{x}_q(t) \right) \\
    \bm{J}_\textrm{T}&= (J_x, J_y)= q\delta\left(  \bm{x}^{\prime}- \bm{x}_q(t)
    \right) \bm{u}_\textrm{T} (\bm{x}_q(t)) \\
    \bm{u}_\textrm{T} &= \left( u_x(\bm{x}_q(t), u_y(\bm{x}_q(t) \right)= \left( \dot{x}_q,
    \dot{y}_q  \right)
  \end{aligned}
\end{gather}

\noindent The value of the integral in Eq. \eqref{eq:mmsr-2-30}, for
given $z$ and $t$, depends on the values of the physical quantities
involved at time $t_q= t\mp (z-z_q(t_q))/c$, from this relation we
can determine $t_q = t_q(z, t)$. We have

\begin{gather}\label{eq:mmsr-2-32}
  \begin{aligned}
    -\fr12 \int_{\it\Omega}  \left(
    \bm{E}_{0\lambda} (\bm{x}^{\prime}_{\textrm{T}})/V_\lambda
    \right)\cdot \bm{J}_\textrm{T} \left( \bm{x}^{\prime}, t\mp
    \fr{z-z^{\prime}}{c} \right) \mathrm{d}^3 x^{\prime} \\
     = - \fr12 q \int_{\it\Omega} \delta(\bm{x}^{\prime}\bm{-x}_q) \left(
    \bm{E}_{0\lambda}(x^{\prime}, y^{\prime})/V_\lambda  \right)\cdot
    \bm{u}_\textrm{T} (\bm{x}_q(t)) \mathrm{d}^3
    x^{\prime}\\
     = - \fr12 q\left( \bm{E}_{0\lambda}\left( x_q(t_q), y_q(t_q)
    \right)/V_\lambda  \right)\cdot \bm{u}_\textrm{T} (\bm{x}_q(t)) \\
     t_q= t\mp \fr{z-z_q(t_q)}{c}
  \end{aligned}
\end{gather}

\noindent If we set $z=z^{\prime}$, correspondingly $z=z_{q}(t_{q})$
for the point charge case, the integral (without the $1/2$) with the
minus sign is the same with the one we know for the auxiliary
current, in the case of a detector with short conductors (no
propagation effects), ~\cite{dris}. Indeed we have

\begin{gather}\label{eq:mmsr-2-33}
 \begin{aligned}
  I_\lambda (t)&= -\int_{\it\Omega} \left(
  \bm{E}_{0\lambda} (\bm{x}^{\prime}_{\textrm{T}})/V_\lambda
  \right)\cdot \bm{J}_\textrm{T} (\bm{x}^{\prime}, t)\mathrm{d}^3x^{\prime}, \quad \lambda =1, 2,
  \ldots \\
          \textrm {and for the point charge, } \\
  I_\lambda (t)&= - q \left( \bm{E}_{0\lambda}\left( x_q(t), y_q(t)
    \right)/V_\lambda  \right)\cdot \bm{u}_\textrm{T} (\bm{x}_q(t)) , \quad \lambda =1, 2, \ldots\\
\end{aligned}
\end{gather}

\noindent It is  clear  that the integrals in Eqs
\eqref{eq:mmsr-2-30}, \eqref{eq:mmsr-2-32}, describe the propagation
with velocity $c$ of currents along the corresponding conductors of
the multi conductor transmission line detector. Namely it is the
propagation of the currents shown in Eqs \eqref{eq:mmsr-2-33} which
are currents induced on the conductors by the moving charges at time
$t=0$ or $t=t_0$ at $z=z_q$.

\noindent We may convince ourselves that this is so by calculating
 the currents from the following known equations %\eqref{eq:mmsr-2-34}%

\begin{equation}
  i_k (z, t)= \oint_k \bm{H}^{(\pm)} \cdot \mathrm{d} \bm{l}
  \label{eq:mmsr-2-34}
\end{equation}

\noindent The line integral, with the appropriate direction, is
taken on a plane with $z= \textrm{constant}$. The path of
integration is along a curve outside the conductor but very close to
it. We remind that the conductor currents are only on the surface
because the conductors are ideal with no resistance.

\noindent So the current at  point $z$ at time $t$ is

\begin{gather}\label{eq:mmsr-2-35}
\begin{aligned}
i_\lambda^{(\pm)}(z, t)&= - \fr{1}{2} \int_{\it\Omega}  \left(
\bm{E}_{0\lambda} (\bm{x}^{\prime}_{\textrm{T}})/V_\lambda
\right)\cdot \bm{J}_\textrm{T} \left( \bm{x}^{\prime}, t \mp
\fr{z-z^{\prime}}{c} \right)\mathrm{d}^3x^{\prime} \\
   \textrm {or} \\
i_\lambda^{(\pm)} (z, t)&= - \fr12 q\left( \bm{E}_{0\lambda}\left(
x_q(t_q), y_q(t_q)
    \right)/V_\lambda  \right)\cdot \bm{u}_\textrm{T}
    (\bm{x}_q(t)) \\
    \\
    t_q&= t\mp \fr{z-z_q(t_q)}{c}
    \end{aligned}
\end{gather}

therefore

\begin{equation}
  i_\lambda ^{(\pm)}(z, t)=
  c \sum_{\lambda ^{\prime}=1 }^{N} c_{\lambda ^{\prime}\lambda }\upsilon^{(\pm)}_{\lambda ^{\prime}} (z,
  t), \quad \lambda =1, 2, \ldots, N
  \label{eq:mmsr-2-36}
\end{equation}

\noindent The $1/2$ is due to the fact that the signal is split into
two signals that move towards opposite directions.

\begin{figure}[htb]
  \centering
 \input{#figures/magnflux.pdf_tex}
  \caption{Magnetic flux ${\it\Psi} _{ij}$ per unit length, through area from i-conductor to 0-conductor, due to the current of conductor
  $j$.} \label{fig:mmsr-2-3}
\end{figure}
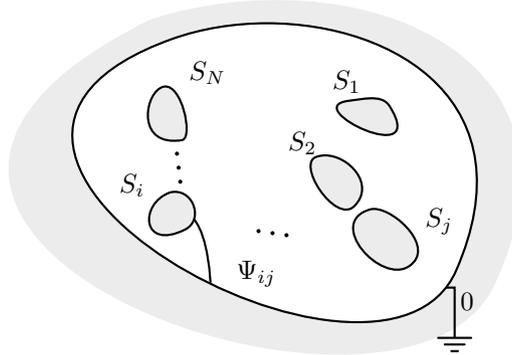

\noindent The conclusion is that the equivalent circuit for the
detector is as in Fig. \ref{fig:mmsr-2-3}.

\noindent It  consists of current  sources each connected to one
conductor of a the multi conductor transmission line. The current
sources are  the same with the ones for small size detectors. This
system is the "internal" equivalent system or the detector system.
The ends of the transmission lines are connected to an external
system.

\noindent Remember that in the small detector size case the current
sources are connected to the internal system  of  capacitances, and
all that  constitutes the "internal" equivalent system or the
detector. This system is connected to the external system.

\noindent In practice, the localization domain of the excitation,
which is between $z_+$, $z_-$, is very small with respect to the
distance $|z-z_+|$ or $|z-z_-|$ from the area we are interested to
know the signal values. It is for this reason that we can accept
that $z^{\prime}\approx z_-\approx z_+\approx z_0=\text{given
constant}$. It can also be assumed that $\bm{x}^{\prime}$, which is
$(\bm{x}^{\prime}_\textrm{T} , z^{\prime})$, is on the given
transverse plane, in position $z^{\prime}=z_0$ at all times. In
other words, we have approximately, $\bm{x}^{\prime}=
(\bm{x}^{\prime}_\textrm{T} , z_0)$ . This is because, in practice,
the wave length is much bigger than $|z_+-z_-|$, so for a given
time, the fields do not change significantly with position
$z^{\prime}$ in the domain of localization of the source. This
simplifies the calculation of the integral and suggests that the
current sources are not distributed, they are localized at one point
along the $z$-axis, the same for all conductors.

\noindent Thus we get

\begin{gather}\label{eq:mmsr-2-37}
\begin{aligned}
  I_\lambda (t)&=- \int_{\it\Omega}  \left(
  \bm{E}_{0\lambda}(x^{\prime}_\textrm{T})/V_\lambda  \right) \cdot \bm{J}_\textrm{T}\left(
  \bm{x}^{\prime}_\textrm{T} , z_0; t
  \right)\mathrm{d}^3x^{\prime}, \quad \lambda =1, 2, \ldots, N \\
  I_\lambda (t)&= - q \left( \bm{E}_{0\lambda}\left( x_q(t), y_q(t)
    \right)/V_\lambda  \right)\cdot \bm{u}_\textrm{T} \left(x_q(t),y_q(t),z_0 \right) , \quad \lambda =1, 2, \ldots\\
\end{aligned}
\end{gather}

\noindent \\

\noindent We will express all the above in matrix form. At first we
described propagation in the two opposite directions without
examining reflections at the two ends. In the matrix formulation it
is easier to include reflections and small losses during the
propagation. We do it by making  use of the well developed theory of
transmission lines in matrix form.

\noindent From the previous relations we know the currents of each
conductor at each point $(z,t)$ and we can determine the
corresponding voltages. If there are not reflections it is easy to
find the signals at the end of the conductors. If there are
reflections things are more complicated as can be seing later on.

\noindent Let us write Eqs. \eqref{eq:mmsr-2-37} in a matrix form,

\begin{gather} \label{eq:mmsr-2-38}
\begin{aligned}
  \left ( i \right )&=
  \begin{bmatrix}
    i_1\\ i_2\\ \vdots \\  i_N
  \end{bmatrix}, \quad  \left ( v \right ) =
  \begin{bmatrix}
    v_1\\ v_2\\ \vdots \\ v_N
  \end{bmatrix}, \quad
  \left ( i \right ) =c[c] \left( v \right )=[Y_{0}]  \left (v  \right ) \\
  [Y_{0}]&=c[c]
  \end{aligned}
\end{gather}

\noindent $[c]$ is the $ ( N \times N ) $  matrix of the capacitance
per unit length coefficients.  Currents and voltages are one column
matrices (column vectors) denoted by $ \left (  \right ) $. All
matrices, denoted by $[$ $]$, are $ ( N \times N ) $ matrices. As we
have seen in Eq. \eqref{eq:mmsr-2-10}, coefficients $c_{ij}$ can be
expressed with respect to the ``usual'' capacitances per unit
length, $C_{lm}$, the two terminal capacitances per unit length.
$[Y_{0}]$ is a conductivity matrix (it is not per unit length). It
is clear that the same relations hold for its matrix elements as is
the case for the matrix elements of $[c]$. Namely $ Y_{0j\lambda }=
Y_{0j\lambda }\leq 0\quad \forall j\neq \lambda , \qquad Y_{0jj}\geq
0 $ .

\noindent By inverting Eqs. \eqref{eq:mmsr-2-38} we find %Eqs
\eqref{eq:mmsr-2-39}%

\begin{equation}
  \left ( v \right ) = \fr{1}{c} [c]^{-1} \left ( i \right ) , \quad \left ( v \right ) =[Z_{0}] \left ( i \right ) , \quad [Z_{0}]=[Y_{0}]^{-1}= \fr{1}{c}
  [c]^{-1}
  \label{eq:mmsr-2-39}
\end{equation}

\noindent We see that to the conductivity matrix, $[Y_{0}]$,
corresponds an impedance matrix $[Z_{0}]$ (not per unit length), it
is the characteristic impedance matrix.For all elements of matrix
$[Z_{0}]$, we have $ Z_{0ij} = Z_{0ji} \geq 0 $.

\noindent Now we will give the (differential) equations that
describe propagation in ideal transmission lines, as the lines we
examined so far are.

\noindent Before doing so we note that, the following equations are
known from multi conductor ideal transmission line theory. In this
case the conductors have not resistance, so currents flow only on
the surfaces of these ideal conductors, as our case is,  and we get
%the following Eqs \eqref{eq:mmsr-2-40}%

\begin{equation}
  [c][L]=\epsilon \mu [E_{\textrm{u}}], \qquad [L]= \epsilon \mu
  [c]^{-1}, \qquad [L]=(1/c^2) [c]^{-1}, \qquad [L][c]= \epsilon \mu /c^{2}
  \label{eq:mmsr-2-40}
\end{equation}

\noindent $[L]$ is the inductance per unit length matrix, it is the
matrix for the external inductances, because there is not current
flow inside the conductors, so there are not internal inductances.
$[E_{\textrm{u}}]$ is the $ ( N\times N ) $ unit matrix, or identity
matrix. We have $ L_{ij}= L_{ji} \geq 0 $ for all matrix elements.

\noindent We can write %Eqs \eqref{eq:mmsr-2-41}%

\begin{gather}\label{eq:mmsr-2-41}
  \begin{aligned}
    \left ( v \right ) &= \fr{1}{c} [c]^{-1} \left ( i \right ) = c[L] \left ( i \right )  \\
    v_{\lambda } ^{(\pm)} (z, t)&= c \sum_{\lambda ^{\prime}=1}^{N} L_{\lambda
    \lambda ^{\prime}} i_{\lambda ^{\prime}}^{(\pm)} (z, t)
  \end{aligned}
\end{gather}

\noindent The physical meaning of the inductance per unit length
coefficients becomes evident from Fig. \ref{fig:mmsr-2-3}. The
current of conductor $j$ creates a magnetic flux ${\it\Psi}_{ij}=
L_{ij} i_j$ per unit length (along $z$, penetrating the area shown
as a line connecting any point on the surface of conductor $i$ with
any point on the surface of enclosing conductor 0. The total
magnetic flux per unit length through that area will be the sum of
all partial fluxes due to the currents of all conductors, i.e. the
following relations %Eq \eqref{eq:mmsr-2-42}% hold

\begin{equation}
{\it\Psi}_i= \sum_{j=1}^{N} {\it\Psi}_{ij} = \sum_{j=1}^{N}
L_{ij}i_j \label{eq:mmsr-2-42}
\end{equation}

\noindent The (differential) equations for transmission lines are
%Eqs \eqref{eq:mmsr-2-43}, below%

\begin{gather} \label{eq:mmsr-2-43}
  \begin{aligned}
    -\pr{i_\lambda }{z}&=  \sum_{\lambda ^{\prime}=1}^{N} c_{\lambda
    ^{\prime}\lambda } \pr{v_{\lambda ^{\prime}}}{t}, \qquad & - \pr{v_\lambda
    }{z}&= \sum_{\lambda ^{\prime}=1}^{N} L_{\lambda ^{\prime}\lambda }
    \pr{i_{\lambda ^{\prime}}}{t} \\
    \text{or } -\pr{}{z} \left ( i \right ) &= [c] \pr{}{t} \left ( v \right ) , \qquad & - \pr{}{z} \left ( v \right ) &= [L]
    \pr{}{t} \left ( i \right )
  \end{aligned}
\end{gather}

\noindent The wave equations for multi conductor transmission lines
are the following %Eqs \eqref{eq:mmsr-2-44}, ones%

\begin{equation}
  \pr{^2i_\lambda }{z^2}= \fr{1}{c^2} \pr{^2i_\lambda }{t^2}, \qquad
  \pr{^2v_\lambda }{z^2}= \fr{1}{c^2} \pr{^2v_\lambda }{t^2}
  \label{eq:mmsr-2-44}
\end{equation}

\noindent These equations, with the proper initial and boundary
conditions, solve the problem of signal propagation over detectors
of many conductors (electrodes). It is clear one can treat the
complicated problem of reflections by imposing the proper boundary
conditions at the ends (terminals) of the detector electrodes.
Remember we know the currents "injected" by the current sources, in
a certain point (given $z$), as functions of time.

\noindent \\

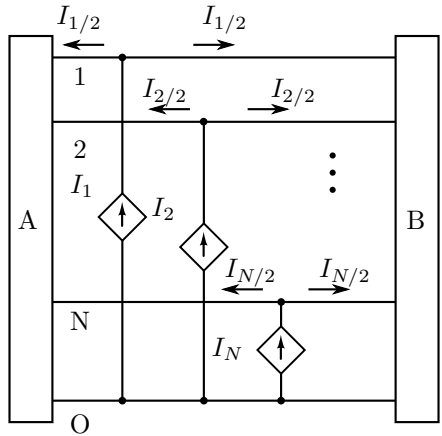
\begin{figure}[!h]
  \centering
 \input{#figures/MCdetequivCIRCUIT1.pdf_tex}
  \caption{Equivalent circuit of multi conductor detector with external (terminal) circuits at both ends.}\label{fig:mmsr-2-4}
\end{figure}

\noindent At this point we summarize what it is proven so far for
multi conductor detectors when the conductors are ideal. One may use
Eqs \eqref{eq:mmsr-2-35},\eqref{eq:mmsr-2-36}, \eqref{eq:mmsr-2-37},
to conclude that any such detector is equivalent to what is depicted
in Fig. \ref{fig:mmsr-2-4}. To each conductor is connected a current
source, at the point along the $z$ direction where the particle hits
the detector.  For each of these current sources the current versus
time is calculated the same way as for (ideal) small size detectors.
Then the propagation is treated by solving the multi conductor
transmission line equations with known terminating networks. We
proved that when the conductors of the ideal transmission line are
excited by the respective current sources, the currents propagate
along the different conductors independent of each other, without
any coupling between them. After the signals reach the terminals,
what happens depends on what is  the connected external circuit. In
general we have reflections and after reflection couplings of the
currents occur. For the voltages propagating along the ideal
conductors there is coupling from the start. The situation is
"reversed" if the excitation is due to voltage sources, which is not
the case for the analysis we follow.

\noindent We comment on a useful result about total charge through
the current source of each conductor, see book by W. Blum, W.
Riegler, L. Rolandi ~\cite{blum}. This result holds for small size
detectors with no propagation effects. Since we proved that the
current sources are the same for long detectors the result holds for
this case too. If a point charge $q$, is moving along a trajectory $
\bm{x} (t) $ from position $ \bm{x}_{0} (t_0) $ to position $
\bm{x}_{1} (t_1) $, the total charge that flows through current
source number  $n$, connected to conductor $n$, is given by %Eq.
\eqref{eq:mmsr-2-45} below%

\begin{gather}\label{eq:mmsr-2-45}
  \begin{aligned}
  Q_{n} = \int_{t_{0}}^{t_{1}} I_{n} \mathrm d t = -  \fr{q}{V_{n}} \int_{t_{0}}^{t_{1}}  \left( \bm{E}_{0\lambda}\left( x_(t), y_(t)
    \right) \right)\cdot \bm{u}_\textrm{T} \left(x_(t),y_(t),z_0 \right) \mathrm d
    t =  \fr{q}{V_{n}} \left ( { \it \Phi} _{0n }(x_1, y_1) - { \it \Phi} _{0n }(x_0,
    y_0) \right )
\end{aligned}
\end{gather}

\noindent We simplified our path description by not including the
index $q$. The charge depends only on the end points of the
trajectory, it does not depend on the specific path. If a pair of
charges $q$, $-q$ are at a point $ \bm{x}_{0} $, where they were
produced, and after some time $q$ moves and arrives at position $
\bm{x}_{1} $ while $ - q $ moves and reaches position $ \bm{x}_{2}
$, the total charge through the current source is given by %the
following Eq. \eqref{eq:mmsr-2-46}%

\begin{gather}\label{eq:mmsr-2-46}
  \begin{aligned}
  Q_{n} = \fr{q}{V_{n}} \left ( { \it \Phi} _{0n }(x_1, y_1) - { \it \Phi} _{0n }(x_2,
    y_2) \right )
\end{aligned}
\end{gather}

\noindent If charge $q$ moves to the surface of conductor (electrode
of the detector) $n$ while charge $-q$ moves to the surface of some
other electrode, the total charge through the source of the $n$
electrode is equal to $q$. When both charges move to other
electrodes, the total charge through the $n$ source is zero. The
conclusion is that, after all charges have arrived at the different
electrodes, the total charge through the source of electrode $n$ is
equal to the charge that has arrived at electrode $n$. From this one
also concludes that the above currents on electrodes that do not
receive any charge are bipolar.

\noindent \\

\noindent In what follows we consider the detector is not an ideal
multi conductor transmission line. There are various "losses", these
include conductor resistance and transverse currents, i.e. currents
in the dielectric which could be not ideal dielectric. They may
exist several different dielectric materials in a cylindrical
(homogeneous along the $z$-axis) arrangement. We assume though, the
situation is such that the currents of the current sources are
calculated the same way as in the ideal case.

\noindent For the propagation the well known matrix treatment of
transmission lines will be used. There are various articles and
books on the subject, like ~\cite{bewley, pipes, pipes1, weeks,
marx, liu, ling, paul}.

\noindent We give formulae relating currents in conductors with
their voltages (potentials) relative to a reference conductor which
has zero  potential, see reference ~\cite{ling}. We have the
following equations %Eqs \eqref{eq:mmsr-2-47},% in matrix form

\begin{gather}\label{eq:mmsr-2-47}
  \begin{aligned}
  \left ( i \right ) =[g] \left ( v \right ) , \qquad g_{ij} \leq 0 \quad \forall i \neq j , g_{ii} \geq
  0
  \end{aligned}
\end{gather}

\noindent $[g]$ is the conductance matrix.

\noindent The above conductances have not direct representation
(realization) with "usual" electric circuit elements. To achieve
that we introduce the direct conductance matrix $G]$,  with elements
$G_{lk}$, which relate currents to voltage differences between
conductors. Each element of direct conductances $R_{lk}$, can be
represented by a resistor $R_{lk}$ connecting node $l$ with node
$k$, and relation $R_{lk}=1/G_{lk}$ holds. Elements $G_{ll}$
correspond to resistors $R_{ll}$ connected between node $l$ and
reference node 0. We have %Eqs \eqref{eq:mmsr-2-48} below%

\begin{gather}\label{eq:mmsr-2-48}
\begin{aligned}
i_{l}&=G_{ll}\upsilon_{l}+\sum_{k=1}^{N}G_{lk}(\upsilon_{l}-\upsilon_{k}),
\quad G_{ll}=\sum_{k=1}^{N}g_{lk}\geq 0, \quad G_{lk}=-g_{lk} \geq 0
\quad \forall l\neq k \\
g_{ll}&=\sum_{k=1}^{N}G_{lk}.
\end{aligned}
\end{gather}

\noindent In what follows, $[g]$ and $[G]$ are conductance matrices,
with elements per unit length, referring to transverse conduction
currents through the dielectric material of the transmission line
and polarization losses in this material.

\noindent Let $[R]$ be the matrix of resistances per unit length for
each conductor of the transmission line. Its elements are positive
and they can take values depending on the approximation used. If the
reference (0th) conductor has no resistance then $[R]$ takes the
simplest form, it is diagonal with the only non zero elements $\quad
R_{11}=r_1, \quad R_{22}=r_2, \quad R_{33}=r_3, ..., R_{NN}=r_{N}$.
In the case the reference conductor could be considered to have
unique resistance, $r_0$  for all currents through the other $N$
conductors, then the elements of matrix [R] are $\quad
R_{11}=r_1+r_0, \quad R_{22}=r_2+r_0, \quad R_{33}=r_3+r_0, ...,
R_{NN}=r_{N}+r_0$,  all the non diagonal elements  are equal to
$r_0$. In the more complicated case, the reference conductor
resistance is different if its current goes through each of the
other conductors, depending on the position of the  conductor
relative to the reference one, one influence of the , proximity
effect. Then we have $\quad R_{11}=r_1+r_{11}, \quad
R_{22}=r_2+r_{22}, \quad R_{33}=r_3+r_{33}, ...,
R_{NN}=r_{N}+r_{NN}$.  All other non diagonal elements $r_{ij}$, in
general, are  different, the matrix could be no symmetric. For the
quasi-TEM analysis to be applicable, the resistances must be
sufficiently small.

\noindent In this case, even if $[L]$ includes both external and
internal inductances, it is assumed the situation is such that its
matrix elements are considered independent of frequency, at least
for the range of frequencies involved. Remember, the internal
inductances exist only if there are currents inside the conductors,
i.e. the conductors have resistance, they are not ideal. If the
conductors are ideal the currents run only across their surfaces
(surface currents) and there exist only external inductances. The
external inductances exist always no matter where the currents
through the conductors are located. The internal inductances and the
resistance of each conductor, depend on the distribution of current
inside the conductor. The distribution depends on the skin depth
which depends on frequency. In general, the conductor resistances
should be low enough so the quasi-TEM approximation holds, this
means that internal inductances might not be so big comparing to
external ones. The resistances will be bigger than the resistances
for dc currents with constant density, all over the (transverse)
cross section of the conductor.

\noindent From multi conductor transmission line theory, we have the
more general than before, relations \eqref{eq:mmsr-2-49} that
follow. Included are losses due to constant conductor resistances
and transverse currents in the surrounding dielectric materials plus
polarization losses, in addition, there might exist several
dielectric media, with homogeneous arrangement along the $z$-axis.
The whole situation is such that quasi-TEM analysis holds.

\begin{gather}\label{eq:mmsr-2-49}
  \begin{aligned}
  -\pr{}{z} \left ( i \right ) &= [g] \left ( v \right ) +[c] \pr{}{t} \left ( v \right ) , \qquad & - \pr{}{z} \left ( v \right ) &=[R] \left ( i \right )  + [L]
    \pr{}{t} \left ( i \right )
  \end{aligned}
\end{gather}

\noindent From the above relations we get formulae for the uncoupled
wave equations, involving only the currents or only the voltages.
They are generalizations of the respective formulae for a two
conductor line. We give the final results

\begin{gather}\label{eq:mmsr-2-50}
  \begin{aligned}
  \pr{^2}{z^2} \left ( i \right ) -[c][L] \pr{^2}{t^2} \left ( i \right ) -\left([c][R]-[g][L] \right )\pr{}{t}
   \left ( i \right ) -[g][R] \left ( i \right ) &=0 \\
  \pr{^2}{z^2} \left ( v \right ) -[c][L] \pr{^2}{t^2} \left ( v \right )
  -\left([c][R]-[g][L] \right )\pr{}{t} \left ( v \right ) -[g][R] \left ( v \right ) &=0
  \end{aligned}
\end{gather}

\noindent  The above two sets of equations can be combined into one
set with column matrices $ ( 2N \times 1 ) $ and square matrices $ (
2N \times 2N ) $.

\noindent In the treatment of common transmission lines, it is
usually considered (with very good accuracy) that $[g]=0$ and
$[G]=0$.

\noindent   We repeat that in what we have done so far, $[c], [L],
[g], [G], [R]$ have constant matrix elements, i.e. they do not
depend on frequency.

\noindent Later we give formulae to analyze more complicated cases,
where there are frequency dependences. Such a case is the influence
of the skin effect on $[R]$ and $[L]$. Even $[c]$ might depend on
frequency. Frequency dependance may exist in the "transverse" losses
through $[g]$.

\noindent For the ideal case with not any type of losses, we have,
$[g]=0$, $[G]=0$ and $[R]=0$ and the above formulae get the
appropriate forms.

\noindent We give, without proof, some relations for the various
matrices, when there is a homogeneous medium between the conductors,
Eqs \eqref{eq:mmsr-2-51}. The conductors are ideal so $[L]$ has
elements the external inductances per unit length.

\begin{gather}\label{eq:mmsr-2-51}
  \begin{aligned}
  [L][c]&=[c][L]=\mu \epsilon [E_{\textrm{u}}]\\
  [L][g]&=[g][L]=\mu \sigma [E_{\textrm{u}}]
  \end{aligned}
\end{gather}

\noindent $\mu$ is the permeability of the homogeneous medium.
$\epsilon$ is the permittivity of the medium and $\sigma$ is its
conductivity.

\noindent In general matrix products do not generally commute so the
proper order of multiplication should be observed.

\noindent The above relations are important because if we find only
one of the matrices, then using the above relations we determine the
other two. It is usually easier to find $[c]$. This can be done
sometimes analytically or by solving electrostatic Maxwell's
equations. We have the following relations %Eqs
 \eqref{eq:mmsr-2-52}%

\begin{gather}\label{eq:mmsr-2-52}
  \begin{aligned}
  [L]&= \mu \epsilon [c]^{-1} \\
  [g]&=(\sigma / \epsilon )  [c]
  \end{aligned}
\end{gather}

\noindent Even when the medium is not homogeneous these matrices are
symmetric and positive-definite. $[L]$ depends on permeability of
the medium. For the materials used $ \mu \approx \mu_0$, which is
the permeability of free space. This means that we may find $[c_0]$
for free space, with $\epsilon = \epsilon_0$ and from relation
$[L]=\mu_0 \epsilon_0 [c_0]^-1$ we may determine $[L]$ even for
inhomogeneous medium. This means that for the same system of ideal
conductors, $[L]$ is the same no matter what the medium is.

\noindent  \\

\noindent At this point it is worth to mention that one way to
derive the above equations for multi conductor transmission lines is
by using the per unit length equivalent circuit. One considers that
the transmission line is represented by a large number of lumped
circuits, each of small length, $\Delta z$,  in comparison to the
wavelengths involved. At the end $\Delta z$ goes to zero and the
number of those circuits goes to infinity. Figure \ref{fig:mmsr-2-5}
shows the general model. Notice that the capacitances shown are
related to the elements of the matrix $[C]$, they are represented by
actual capacitors. The conductances are the realizable ones, related
to $[G]$, they are represented by actual resistors each equal to
$1/(G_{ij} \Delta z)$. They are related to the elements of $[c]$ and
$[g]$, that appear in the transmission line equations. We assume the
reference conductor has a single resistance per unit length$r_0$. We
repeat, all elements of the above matrices are independent of
frequency. From those lumped circuits, by going to the limit we
mentioned,  one can deduce the formulae for transmission lines we
gave so far.

\begin{figure}[!h]
  \centering
\input{#figures/LUMPEDcirctransmline.pdf_tex}
  \caption{The per unit length circuit model for a multi conductor transmission line.} \label{fig:mmsr-2-5}
\end{figure}
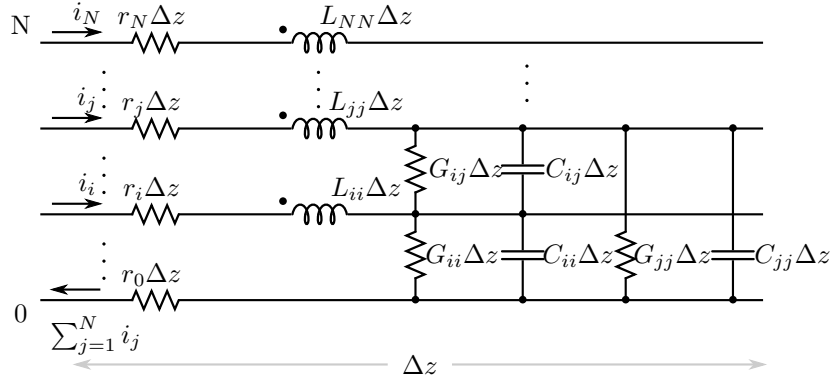

\noindent Among the various ways to find solutions of the above
equations, is by taking Laplace or Fourier transforms with respect
to time. The two are related by exchanging the real frequency,
$\omega$, with the complex one, $s$, according to $j \omega
\leftrightarrow s$.

\noindent We assume that initially, at $t=0$, the lines have zero
voltages and currents, then by laplace transforming with respect to
time, Eqs \eqref{eq:mmsr-2-47} lead to the following relations %Eqs
\eqref{eq:mmsr-2-53}%

\begin{gather}\label{eq:mmsr-2-53}
  \begin{aligned}
   \drac { \left ( I(z,s) \right ) } z &= - \left ([g]+s[c] \right ) \left ( V(z,s) \right )  \\
   \drac { \left ( V(z,s) \right ) } z &= - \left ( [R]
  +s[L] \right ) \left ( I(z,s) \right )
  \end{aligned}
\end{gather}

\noindent We may proceed from Eqs \eqref{eq:mmsr-2-53}, or from
\eqref{eq:mmsr-2-50} to get %Eqs \eqref{eq:mmsr-2-54} below%

\begin{gather} \label{eq:mmsr-2-54}
  \begin{aligned}
   \drac {^2  \left ( I(z,s) \right ) } {z^2} &= \left ( [g]+s[c] \right ) \left ([R]+s[L] \right )  \left ( V(z,s) \right ) \\
   \drac {^2  \left ( V(z,s) \right ) } {z^2} &= \left ( [R]
  +s[L] \right )\left ( [g]+s[c] \right ) \left ( I(z,s) \right )
  \end{aligned}
\end{gather}

\noindent These formulae hold even when some or all above matrices
have elements that depend on frequency. We do not go into details
about such dependences, among other references, one could consult
the book by C. R. Paul, Ref.  ~\cite{paul}.

\noindent Another form of the above equations is %given below, Eqs
\eqref{eq:mmsr-2-55}%

\begin{gather} \label{eq:mmsr-2-55}
  \begin{aligned}
   [Z(s)]&=[R]+s[L],  [Y(s)]=[g]+s[c] \\
   \drac { \left ( I(z,s) \right ) } z &= - [Y(s)] \left ( V(z,s) \right ) \\
   \drac { \left ( V(z,s) \right ) } z &= - [Z(s)]  \left ( I(z,s) \right ) \\
  \drac {^2  \left ( I(z,s) \right ) } {z^2} &=  [Y(s)][Z(s)]  \left ( I(z,s) \right ) \\
  \drac {^2  \left ( V(z,s) \right ) } {z^2} &=  [Z(s)][Y(s)] \left ( V(z,s) \right )
  \end{aligned}
\end{gather}

\noindent Usually, $[R], [L]$ are taken to depend on frequency,
while $[c], [g]$ are constant. $[Z]$ is the per unit length
impedance matrix and $[Y]$ the per unit length admittance matrix.
One has to solve the above equations imposing the proper terminal
constrains, that depend on the networks connected to the terminals
of the multi conductor line.

\noindent The general case is difficult to treat. We only mention
that there is attenuation along the line as it is true for (1+1)
lines. As we said before, in general, if there are more than one
dielectric material in a homogeneous arrangement  along the
$z$-direction, or if the conductors are not ideal, they have
resistances, they might be $N$ (maximum) number of propagation modes
each one with different speed.

\noindent The generalization of the characteristic impedance is the
impedance $( N \times N )$ matrix which is in general frequency
dependent. Reflections occur at both terminals if not properly
terminated.

\noindent For signals propagating along only one dimension, either
$+z$ or $-z$, we have for the solutions the form %shown below in Eqs
\eqref{eq:mmsr-2-56}%

\begin{gather}\label{eq:mmsr-2-56}
  \begin{aligned}
  \left ( I(z,s) \right ) = \left ( I_{0}(s) \right )  \exp  \left ( \pm \gamma (s) z \right ) , \quad
  \left ( V(z,s) \right ) = \left ( V_{0}(s) \right )  \exp  \left (\pm \gamma (s) z
  \right )  \\
   \det [[Z(s)][Y(s)]-\gamma ^2 (s)   [E_{\textrm{u}}]]=0, \quad
   \det [[Y(s)][Z(s)]-\gamma ^2 (s)  [E_{\textrm{u}}]]=0 \\
   \gamma _{j} (s), \quad j =1, 2, \ldots, N
  \end{aligned}
\end{gather}

\noindent The - sign in the exponentials correspond to propagation
along the $+z$ direction and the + sign along the $-z$ direction.
The relations in the second line determine the propagation
coefficient $ \gamma (s) $, from witch one could determine the
signal attenuation and the various speeds of propagation. Both
relations with the zeroing of determinants give the same results. In
some cases the propagation coefficients take the simple form $a+bs$,
with $a$ and $b$ positive constants. In this case it is easy to
determine the attenuation and speed of propagation. As we said
before, in general, for multi conductor transmission lines, there
are (maximum) $N$ different modes of propagation, $N$ different $
\gamma $, $ \gamma _1, \gamma _2, ... \gamma_N $. It is taken Real$
\gamma_{j} \geq 0$ and Im$ \gamma_{j} \geq 0 $. From these $N$
different speeds, $v_1(s), v_2(s), ... v_{N}(s)$ could be
calculated. One determines the signals as functions of position and
time by taking the inverse Laplace transform of the above relations.
Each solution is superposition of all modes of propagation with all
different propagation coefficients. The appropriate boundary
conditions at the sources and terminals should be taken into
account.

\noindent At this point we analyze mainly the case of lossless multi
conductor lines, in general, with many dielectric materials
distributed as mentioned before.

\noindent We make $[R]=0$ and $[g]=0$, so Eqs \eqref{eq:mmsr-2-55} ,
become %Eqs \eqref{eq:mmsr-2-57}%

\begin{gather}\label{eq:mmsr-2-57}
  \begin{aligned}
   [Z(s)]&=s[L],  [Y(s)]=s[c] \\
   \drac { \left ( I(z,s) \right ) } z &=  -s[c] \left ( V(z,s) \right )  \\
   \drac { \left ( V(z,s) \right ) } z &= - s[L] \left ( I(z,s) \right )  \\
  \drac {^2  \left ( I(z,s) \right ) } {z^2} &= s^2 [c][L] \left ( I(z,s) \right )  \\
  \drac {^2  \left ( V(z,s) \right ) } {z^2} &= s^2 [L][c] \left ( V(z,s) \right )
  \end{aligned}
\end{gather}

\noindent Since $[R]=0$ then $[L]$ is constant and equal to the
external inductance matrix. $[c]$ is constant too. It is easy to see
that for a homogeneous medium there is only one propagation
coefficient because the solutions for the values for the propagation
coefficients are degenerate and have a single value. The final
relation for the single speed is %as in the following Eqs
 \eqref{eq:mmsr-2-58}%

\begin{gather}\label{eq:mmsr-2-58}
  \begin{aligned}
  [L][c] = [c][L] = (1/c^2) [E_{ \textrm{u}}],  \quad  c = v
  \end{aligned}
\end{gather}

\noindent One might want to examine the problem where, Eq.
 \eqref{eq:mmsr-2-58} is not true, this could be the case if
 $[L]$ and $[c]$ are not related to each other by Eqs
 \eqref{eq:mmsr-2-52}, because there might be several dielectrics
present and in addition one might include in $[L]$ the internal
(small) inductances as constants. One still ignores $[R]$.

\noindent For the single propagation constant we have %Eq.
 \eqref{eq:mmsr-2-59}%

\begin{gather}\label{eq:mmsr-2-59}
  \begin{aligned}
  \gamma = s/c
  \end{aligned}
\end{gather}

\noindent The traveling signals in both directions are given by relations %Eqs
 \eqref{eq:mmsr-2-60}%

\begin{gather}\label{eq:mmsr-2-60}
  \begin{aligned}
  \left ( I(z,s) \right ) &=  \left ( I_{+}(s) \right )  \exp  \left ( - \gamma
 z  \right ) +  \left ( I_{-}(s)  \right )  \exp  \left ( + \gamma z \right ) \\
  \left ( V(z,s) \right ) &= [Z_{0}]  \left ( \left ( I_{+}(s) \right
 ) \exp  \left ( - \gamma z \right ) - \left ( I_{-}(s) \right )  \exp
  \left ( + \gamma z  \right )
 \right ) \\
  \end{aligned}
\end{gather}

\noindent $I_{+} $ has as matrix elements  the amplitudes of the
current  signals traveling in the $+z$ direction. $I_{-} $ the ones
traveling in the $-z$ direction. They can be estimated by imposing
the proper boundary condition, exciting currents and terminal
networks. $[Z_{0}]$ is the characteristic impedance matrix which is
not per unit length (we have seing that before, Eqs
\eqref{eq:mmsr-2-39}). $[Y_{0}]$ is the corresponding characteristic
conductance matrix. The following equations relations %Eq.
\eqref{eq:mmsr-2-61}% hold

\begin{gather}\label{eq:mmsr-2-61}
  \begin{aligned}
  [Z_{0}(s)] = [Y]^{-1} \sqrt {[Y][Z]}, \quad [Y_{0}]=[Z_{0}]^{-1}
  \end{aligned}
\end{gather}

\noindent It easy to see that for (1+1) conductors line we get the
simple result we know.

\noindent Let us assume we have the situation of equations %Eqs
 \eqref{eq:mmsr-2-62} below%

\begin{gather}\label{eq:mmsr-2-62}
  \begin{aligned}
  [Y]&=s[c], \quad  [Z]=s[L] \\
  [Z_{0}(s)] &= [c]^{-1} \sqrt{[c][L]}
  \end{aligned}
\end{gather}

\noindent If we assume relation between $[L]$ and $[c]$ holds as in
Eq. \eqref{eq:mmsr-2-52} then we get %Eq. \eqref{eq:mmsr-2-63}%

\begin{gather}\label{eq:mmsr-2-63}
  \begin{aligned}
  [Z_{0}]= \fr {1}{c} [c]^{-1}
  \end{aligned}
\end{gather}

\noindent This is the same result as in Eq. \eqref{eq:mmsr-2-39}.

\noindent At this point we give some general formulae for treating a
single reflection at one terminal, see  ~\cite{paul}. The general
formula for the voltage reflection matrix at the load is, $ [ \Gamma
_{V \textrm {L} } (s)] $. The current reflection matrix is $ [
\Gamma _{I \textrm {L} } (s)] $. They are the generalizations of the
reflection coefficients for usual (1+1) transmission lines. For the
reflection matrix at the source $ \textrm {L} \rightarrow \textrm
{S} $. We have the formula %shown in Eqs \eqref{eq:mmsr-2-64}%

\begin{gather}\label{eq:mmsr-2-64}
  \begin{aligned}
  [ \Gamma_{V \textrm {L} }(s) ] &=  \left ( [Z_{ \textrm {L} }] - [Z_{0}] \right ) \left ( [Z_{ \textrm {L} }] + [Z_{0}] \right )
  ^{-1} \\
  [ \Gamma_{I \textrm {L} }(s) ] &= - [Y_{0}]  [ \Gamma_{V \textrm {L}
  }(s)][Z_{0}]
  \end{aligned}
\end{gather}

\noindent $ [Z_{ \textrm {L} }] $ is the matrix form of the network
connected at the terminal of the line, which has $ N $ inputs.

\noindent Let $ \left ( V_{ \textrm {f} } \right ) $ the column of
signal complex voltages that travel along the $+z$ (forward)
direction towards the load $[[Z_{ \textrm {L} }]$ at the terminal of
the line at point $z=l$. $ \left ( I_{ \textrm {f} } \right ) $ is
the corresponding column of the complex current signals. Signals $
\left ( V_{ \textrm {b} } \right ) $  and $ \left ( I_{ \textrm {b}
} \right ) $ are the corresponding reflected signals. The following
relations% in Eq. \eqref{eq:mmsr-2-65}% hold

\begin{gather}\label{eq:mmsr-2-65}
  \begin{aligned}
  \left ( V_{ \textrm {b} } \right ) &= [ \Gamma_{V \textrm {L} }(s) ] \left ( V_{ \textrm {f} } \right
  ) \\
  \left ( I_{ \textrm {b} } \right ) &= [ \Gamma_{I \textrm {L} }(s) ]
  \left ( I_{ \textrm {f} } \right
  )
  \end{aligned}
\end{gather}

\noindent It is clear that, if for the reflection matrices, relation
$ [ \Gamma] = 0 $ holds, then there are not reflections. The
reflection matrices are zero if the matrix of the terminal network
is the same with the characteristic impedance matrix, i.e. $ [Z_{
\textrm {L} }] = [Z_{0}] $ . This is the impedance matching
condition, matched line.

\noindent For signals traveling in the $+z$, forward and $-z$,
backward directions we have the formulae %\eqref{eq:mmsr-2-66} below%

\begin{gather}\label{eq:mmsr-2-66}
  \begin{aligned}
  \left ( V_{ \textrm {f} } \right ) &= [Z_{0}]  \left ( I_{ \textrm {f} } \right )
   , \quad  \left ( I_{ \textrm {f} } \right ) = [Y_{0}]  \left ( V_{
     \textrm {f} } \right )
  ) \\
  \left ( V_{ \textrm {b} } \right ) &= -[Z_{0}]  \left ( I_{ \textrm {b} } \right
  ), \quad \left ( I_{ \textrm {b} } \right ) = - [Y_{0}]  \left ( V_{
     \textrm {b} } \right )
  \end{aligned}
\end{gather}

\noindent Another interesting concept is the transmission or
transfer matrix. It is the generalization of the corresponding
transfer coefficient for (1+1) lines. This matrix is $
 \left ( [E_{\textrm{u}}] + [ \Gamma ] \right )  $ The definition is based on the fact
that, if we consider the "incoming" and reflected signals at the
terminal, the voltage or current signal $ [S_{ \textrm{t}}] $ that
the terminal "sees" is the sum of the two. The same is true for the
currents. So we have %the following Eq. \eqref{eq:mmsr-2-67}%

\begin{gather}\label{eq:mmsr-2-67}
  \begin{aligned}
  \left ( S_{ \textrm {t} } \right ) = \left ( [E_{\textrm{u}}] + [ \Gamma ] \right ) \left ( S_{ \textrm
  {t}} \right )
  \end{aligned}
\end{gather}

\noindent The problem of termination of the  multi electrode long
detectors is a complicated procedure.  Let us first generalize Eqs
\eqref{eq:mmsr-2-47} and \eqref{eq:mmsr-2-48} to hold for complex
values too. We refer to signals propagating in the $+z$ direction
when they reach the terminal. We may examine reflections by let us
assume that the line is matched and no reflections exist. We use Eqs
\eqref{eq:mmsr-2-66} and we get %Eqs \eqref{eq:mmsr-2-68} below%

\begin{gather}\label{eq:mmsr-2-68}
  \begin{aligned}
   \left ( I_{ \textrm {t}} \right ) &= \left ( I_{ \textrm {f}} \right
   ) = \left ( I \right ) \\
   \left ( V_{ \textrm {t}} \right ) &= \left ( V_{ \textrm {f}} \right
   ) = \left ( V \right ) \\
  \left ( I \right ) &=[Y_{0}] \left ( V \right ) \\
   I_{l}&=G_{ll}V_{l}+\sum_{k=1}^{N}G_{lk}(V_{l}-V_{k}),
 \quad G_{ll}=\sum_{k=1}^{N}Y_{0lk}, \quad G_{lk}=-Y_{0lk}
\quad \forall l\neq k \\
Y_{0ll}&=\sum_{k=1}^{N}G_{lk}
  \end{aligned}
\end{gather}

\noindent We expressed the currents in terms of the voltage
differences between conductors, so we could have a realization of
the terminal network with usual electric components. Let us assume
we have the ideal case where $[Y_{0}]=c[c]$. This means all matrix
elements are real. In this case the $[G]$ matrix has (real) non
negative elements.

\noindent  Eqs \eqref{eq:mmsr-2-66} leads us to an equivalent to a
circuit in which each conduct is represented by a node, and the
distinct nodes $l, k$ are connected by a resistor $R_{lk}$. The node
$l$ is also connected to the ground by a resistor $R_{ll}$. These
resistors are the elements of a (symmetric) matrix $[R]$ with the
elements $R_{lk}=G_{lk}^{-1}$.

\noindent To have the multi conductor transmission line matched
(exactly terminated) at both ends, we have to connect to its ends a
complicated network of resistors indicated by the above matrix of
the direct conductances.

\noindent For  the  case  of a 2+1 conductor line, see reference
~\cite{marx},  one has the simple 3 resistor  circuit of Fig.
\ref{fig:mmsr-2-6}. This is accomplished following  the  above
procedure, using the conductivity matrix $[Y_{0}]$. We have %Eqs
 \eqref{eq:mmsr-2-69} below%

\begin{gather}\label{eq:mmsr-2-69}
\begin{aligned}
  G_{11}&=Y_{011}+Y_{012}, \quad \ G_{22}=Y_{012}+Y_{022} \\
  G_{12}&=G_{21}=-Y_{012}, \quad \ R_{11}=1/(Y_{011}+Y_{012})\\
  R_{22}&=1/(Y_{022}+Y_{012}), \quad \ R_{12}=-1/Y_{012}.
\end{aligned}
\end{gather}

\noindent The circuit of Fig. \ref{fig:mmsr-2-7} is a different one
but equivalent to it. For  this  case the resistors are  calculated
more easily from the impedance matrix $[Z]$, we have the following
equations % Eqs  \eqref{eq:mmsr-2-70}%

\begin{equation}
  R_{1}=Z_{011}-Z_{012}, \quad \ R_{2}=Z_{022}-Z_{012}, \quad \ R_{0}=Z_{012}.
  \label{eq:mmsr-2-70}
\end{equation}

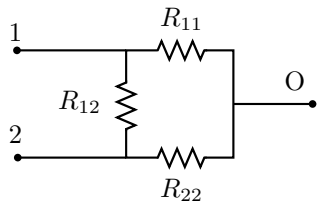
\begin{figure}[!h]
  \centering
 \input{#figures/termincirc1.pdf_tex}
  \caption{An exact terminating resistor circuit.}\label{fig:mmsr-2-5}
\end{figure}

\begin{figure}[!h]
  \centering
 \input{#figures/termincirc2.pdf_tex}
  \caption{Another exact terminating resistor circuit .}\label{fig:mmsr-2-6}
\end{figure}
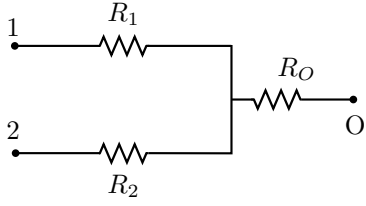

\noindent For $N$ electrodes (the reference conductor not included)
one needs $N(N+1)/2$ resistors! If one has three internal electrodes
surrounded by the grounded electrode, one needs 6 appropriate
resistors on each end. Amari and Bornemann, ~\cite{amari}, examine a
case where there are electrodes of the strip type on a plane
parallel to each other. They achieve a so called optimum termination
(for lowest input return loss), by using resistors interconnecting
only the neighboring electrodes and resistors from each electrode to
the ground. For 10 electrodes exact termination needs 55 appropriate
resistors. If one could apply the mentioned above  approximate
method, 19 appropriate resistors are needed.

\noindent At the end we should say a few things on the need or not
of the use of the transmission line description. A rule of thumb is
that, in general, if for the length, $l$, of a circuit and the
wavelength $\lambda$ of waves involved the relation $l \geq \lambda$
then the propagation effects are important and reflections might
lead to behavior different from the expected one for small size
circuits. For small size circuits even though one may argue that
there are propagation effects, they are such that not deviations
exist from the usual description without propagation effects. In the
case of detectors we can understand that the need of transmission
line description taking into account reflections at the ends of a
detector, could have a significant effect, depending on the detector
length in relation to the wave lengths of the induced signals and on
the (electronic) shaping of the pulses at the readout, which
influences the wave lengths involved in the "final", after shaping
signals.  Our analysis is needed if the "final" wave lengths are
sufficiently small in comparison to the detector length.

\section{Example: Long cylindrical detector of circular cross-section with a wire
along its axis}

\noindent In this case, the position variables are the known
variables of cylindrical coordinates $\varphi , r$. The external
conductor is assumed to have zero potential ($S_0$). We will examine
the case of a point ion charge, whose drift velocity is given by
relation $u=\mu E$, $\mu$=constant  and the motion of the charge is
radial on the transverse [to the $z-$axis] plane, in position
$z_q=\text{constant}$. The charge starts from initial position
$r_0$. Obviously $\varphi =\text{constant}$, independent of time. It
is easy to get the following relations %Eqs \eqref{eq:mmsr-ex2-1}%
(see, ~\cite{radeka,dris,blum})

\begin{gather} \label{eq:mmsr-ex2-1}
\begin{aligned}
  E&=E_ \textrm{T}=\fr{V_a}{\ln(b/a)} \fr{1}{r} \\
  u_\textrm{T} &= \mu  \fr{V_a}{\ln(b/a)} \fr{1}{r}, \qquad r^2= r_0^2+2\mu
  \fr{V_a}{\ln(b/a)}t, \qquad r^2=r_0^2 \left( 1+\fr{t}{t_0} \right) \\
  t_0&= \fr{r_0^2}{2\mu V_a} \ln\left( \fr{b}{a} \right)
\end{aligned}
\end{gather}

\noindent We assume that (high) voltage, i.e. the detector bias that
moves the charges is constant, $V_a$, so from Eqs.
\eqref{eq:mmsr-2-32} and \eqref{eq:mmsr-2-44} for the current and
voltage ``pulse'' we find %Eqs \eqref{eq:mmsr-ex2-2}%

\begin{gather}
  \begin{aligned}
    i (\bm{x}, t) &= -\fr12 \fr{1}{2\pi } \fr{\sqrt{\mu / \epsilon }q\mu
    V_a}{r_0^2\ln(b/a)} \fr{1}{\left[ \left( 1+\fr{t}{t_0} \right)\mp
    \fr{z-z_q}{ct_0} \right]} \\
    v (\bm{x}, t) &= -\fr12 \fr{q\mu V_a}{r_0^2\left(\ln(b/a)\right)^2}
    \fr{1}{\left[ \left( 1+\fr{t}{t_0} \right)\mp \fr{z-z_q}{ct_0}
    \right]} \label{eq:mmsr-ex2-2} \\
    Z_0&= \fr{v(z, t)}{i(z, t)}= \fr{1}{2\pi } \sqrt{\dfrac{\mu }{\epsilon}}
    \ln \left( \frac{b}{a} \right)=\text{characteristic resistance of the transmission line}
  \end{aligned}
\end{gather}
\\

%% file: figures/figures/transvsys.pdf_tex
%% Creator: Inkscape 0.48.3.1, www.inkscape.org
%% Accompanies image file 'transvsys.pdf' (pdf, eps, ps)
\begingroup%
  \makeatletter%
  \providecommand\color[2][]{%
    \errmessage{(Inkscape) Color is used for the text in Inkscape, but the package 'color.sty' is not loaded}%
    \renewcommand\color[2][]{}%
  }%
  \providecommand\transparent[1]{%
    \errmessage{(Inkscape) Transparency is used (non-zero) for the text in Inkscape, but the package 'transparent.sty' is not loaded}%
    \renewcommand\transparent[1]{}%
  }%
  \providecommand\rotatebox[2]{#2}%
  \ifx\svgwidth\undefined%
    \setlength{\unitlength}{212.17565918bp}%
    \ifx\svgscale\undefined%
      \relax%
    \else%
      \setlength{\unitlength}{\unitlength * \real{\svgscale}}%
    \fi%
  \else%
    \setlength{\unitlength}{\svgwidth}%
  \fi%
  \global\let\svgwidth\undefined%
  \global\let\svgscale\undefined%
  \makeatother%
  \begin{picture}(1,0.61404505)%
    \put(0,0){\includegraphics[width=\unitlength]{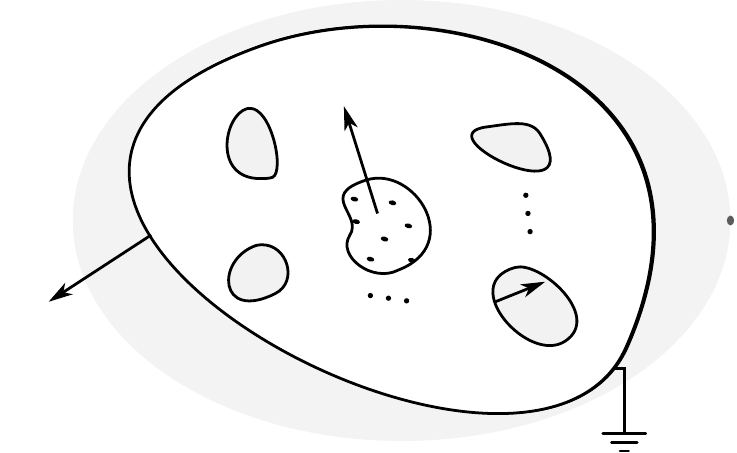}}%
    \put(0.25873072,0.43312348){\color[rgb]{0,0,0}\makebox(0,0)[b]{\smash{$S_1$}}}%
    \put(0.95249553,0.32000965){\color[rgb]{0,0,0}\makebox(0,0)[b]{\smash{$S_0$}}}%
    \put(0.80921802,0.24460044){\color[rgb]{0,0,0}\makebox(0,0)[b]{\smash{$S_\ell$}}}%
    \put(0.40200826,0.16919122){\color[rgb]{0,0,0}\makebox(0,0)[b]{\smash{$S_N$}}}%
    \put(0.40200826,0.16919122){\color[rgb]{0,0,0}\makebox(0,0)[b]{\smash{$S_N$}}}%
    \put(0.38692641,0.32000965){\color[rgb]{0,0,0}\makebox(0,0)[b]{\smash{$\rho(\bm{x},t)$}}}%
    \put(0.50758116,0.47836901){\color[rgb]{0,0,0}\makebox(0,0)[b]{\smash{$\bm{u}(\bm{x})$}}}%
    \put(0.06675814,0.17370138){\color[rgb]{0,0,0}\makebox(0,0)[b]{\smash{$\bm{n}$}}}%
    \put(0.63232728,0.18878322){\color[rgb]{0,0,0}\makebox(0,0)[b]{\smash{$\bm{n}$}}}%
    \put(0.6533381,0.10097045){\color[rgb]{0,0,0}\makebox(0,0)[b]{\smash{$\mathit{\Omega}$}}}%
    \put(0.78659526,0.43312348){\color[rgb]{0,0,0}\makebox(0,0)[b]{\smash{$S_2$}}}%
    \put(0.86922281,0.07897744){\color[rgb]{0,0,0}\makebox(0,0)[b]{\smash{0}}}%
  \end{picture}%
\endgroup%

%% file: figures/figures/longitudinal.pdf_tex
%% Creator: Inkscape 0.48.3.1, www.inkscape.org
%% Accompanies image file 'longitudinal.pdf' (pdf, eps, ps)
\begingroup%
  \makeatletter%
  \providecommand\color[2][]{%
    \errmessage{(Inkscape) Color is used for the text in Inkscape, but the package 'color.sty' is not loaded}%
    \renewcommand\color[2][]{}%
  }%
  \providecommand\transparent[1]{%
    \errmessage{(Inkscape) Transparency is used (non-zero) for the text in Inkscape, but the package 'transparent.sty' is not loaded}%
    \renewcommand\transparent[1]{}%
  }%
  \providecommand\rotatebox[2]{#2}%
  \ifx\svgwidth\undefined%
    \setlength{\unitlength}{228.21594238bp}%
    \ifx\svgscale\undefined%
      \relax%
    \else%
      \setlength{\unitlength}{\unitlength * \real{\svgscale}}%
    \fi%
  \else%
    \setlength{\unitlength}{\svgwidth}%
  \fi%
  \global\let\svgwidth\undefined%
  \global\let\svgscale\undefined%
  \makeatother%
  \begin{picture}(1,0.34453474)%
    \put(0,0){\includegraphics[width=\unitlength]{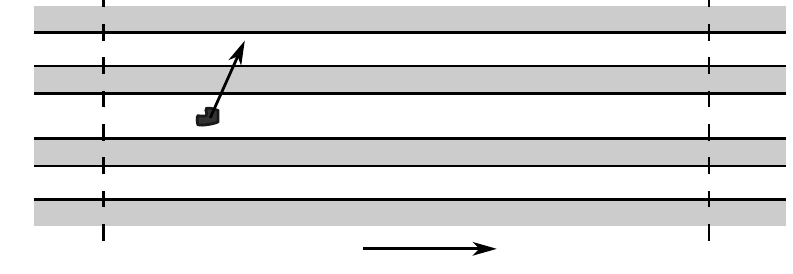}}%
    \put(0.36129855,0.17850117){\color[rgb]{0,0,0}\makebox(0,0)[b]{\smash{$\rho(\bm{x}, t)$}}}%
    \put(0.39635307,0.27665382){\color[rgb]{0,0,0}\makebox(0,0)[b]{\smash{$u(\bm{x})$}}}%
    \put(0.00785646,0.30965303){\color[rgb]{0,0,0}\makebox(0,0)[b]{\smash{0}}}%
    \put(0.00785646,0.23234053){\color[rgb]{0,0,0}\makebox(0,0)[b]{\smash{1}}}%
    \put(0.00785646,0.14119878){\color[rgb]{0,0,0}\makebox(0,0)[b]{\smash{2}}}%
    \put(0.00785646,0.06427906){\color[rgb]{0,0,0}\makebox(0,0)[b]{\smash{0}}}%
    \put(0.4130827,0.02026076){\color[rgb]{0,0,0}\makebox(0,0)[b]{\smash{$z$}}}%
    \put(0.13284266,0.00626976){\color[rgb]{0,0,0}\makebox(0,0)[b]{\smash{$S_-$}}}%
    \put(0.89703115,0.00626976){\color[rgb]{0,0,0}\makebox(0,0)[b]{\smash{$S_+$}}}%
  \end{picture}%
\endgroup%

%% file: figures/magnflux.pdf_tex
%% Creator: Inkscape 0.48.3.1, www.inkscape.org
%% Accompanies image file 'magnflux.pdf' (pdf, eps, ps)
\begingroup%
  \makeatletter%
  \providecommand\color[2][]{%
    \errmessage{(Inkscape) Color is used for the text in Inkscape, but the package 'color.sty' is not loaded}%
    \renewcommand\color[2][]{}%
  }%
  \providecommand\transparent[1]{%
    \errmessage{(Inkscape) Transparency is used (non-zero) for the text in Inkscape, but the package 'transparent.sty' is not loaded}%
    \renewcommand\transparent[1]{}%
  }%
  \providecommand\rotatebox[2]{#2}%
  \ifx\svgwidth\undefined%
    \setlength{\unitlength}{212.17565918bp}%
    \ifx\svgscale\undefined%
      \relax%
    \else%
      \setlength{\unitlength}{\unitlength * \real{\svgscale}}%
    \fi%
  \else%
    \setlength{\unitlength}{\svgwidth}%
  \fi%
  \global\let\svgwidth\undefined%
  \global\let\svgscale\undefined%
  \makeatother%
  \begin{picture}(1,0.61404505)%
    \put(0,0){\includegraphics[width=\unitlength]{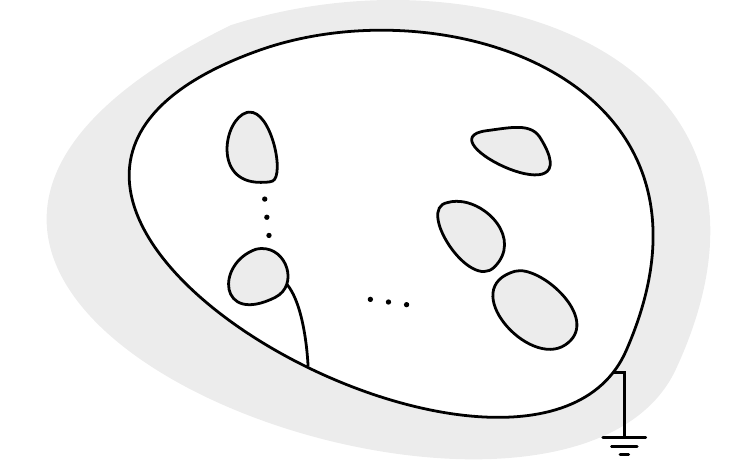}}%
    \put(0.65802281,0.46723387){\color[rgb]{0,0,0}\makebox(0,0)[b]{\smash{$S_1$}}}%
    \put(0.82019899,0.22441519){\color[rgb]{0,0,0}\makebox(0,0)[b]{\smash{$S_j$}}}%
    \put(0.28061543,0.28556781){\color[rgb]{0,0,0}\makebox(0,0)[b]{\smash{$S_i$}}}%
    \put(0.41304397,0.48220415){\color[rgb]{0,0,0}\makebox(0,0)[b]{\smash{$S_N$}}}%
    \put(0.49889545,0.13360468){\color[rgb]{0,0,0}\makebox(0,0)[b]{\smash{$\Psi_{ij}$}}}%
    \put(0.57992647,0.35888324){\color[rgb]{0,0,0}\makebox(0,0)[b]{\smash{$S_2$}}}%
    \put(0.86922281,0.07897744){\color[rgb]{0,0,0}\makebox(0,0)[b]{\smash{0}}}%
  \end{picture}%
\endgroup%

%% file: figures/figures/MCdetequivCIRCUIT1.pdf_tex
%% Creator: Inkscape 0.48.3.1, www.inkscape.org
%% Accompanies image file 'MCdetequivCIRCUIT1.pdf' (pdf, eps, ps)
\begingroup%
  \makeatletter%
  \providecommand\color[2][]{%
    \errmessage{(Inkscape) Color is used for the text in Inkscape, but the package 'color.sty' is not loaded}%
    \renewcommand\color[2][]{}%
  }%
  \providecommand\transparent[1]{%
    \errmessage{(Inkscape) Transparency is used (non-zero) for the text in Inkscape, but the package 'transparent.sty' is not loaded}%
    \renewcommand\transparent[1]{}%
  }%
  \providecommand\rotatebox[2]{#2}%
  \ifx\svgwidth\undefined%
    \setlength{\unitlength}{160.80001221bp}%
    \ifx\svgscale\undefined%
      \relax%
    \else%
      \setlength{\unitlength}{\unitlength * \real{\svgscale}}%
    \fi%
  \else%
    \setlength{\unitlength}{\svgwidth}%
  \fi%
  \global\let\svgwidth\undefined%
  \global\let\svgscale\undefined%
  \makeatother%
  \begin{picture}(1,0.99828509)%
    \put(0,0){\includegraphics[width=\unitlength]{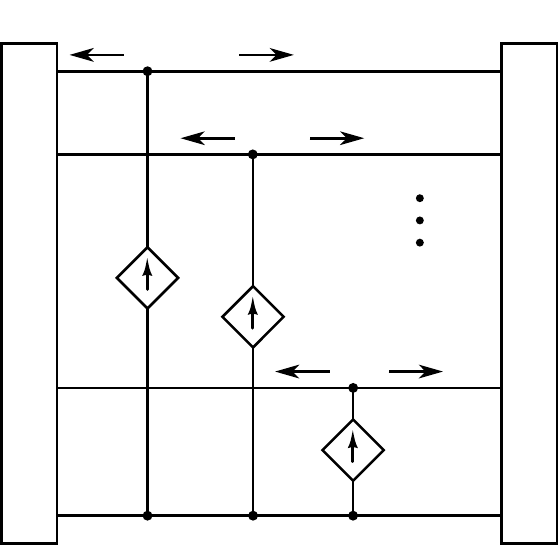}}%
    \put(0.04532315,0.47266594){\color[rgb]{0,0,0}\makebox(0,0)[b]{\smash{A}}}%
    \put(0.94544092,0.47266594){\color[rgb]{0,0,0}\makebox(0,0)[b]{\smash{B}}}%
    \put(0.16642094,0.00077343){\color[rgb]{0,0,0}\makebox(0,0)[b]{\smash{O}}}%
    \put(0.16642094,0.23957921){\color[rgb]{0,0,0}\makebox(0,0)[b]{\smash{N}}}%
    \put(0.16642094,0.63758912){\color[rgb]{0,0,0}\makebox(0,0)[b]{\smash{2}}}%
    \put(0.16642094,0.80674333){\color[rgb]{0,0,0}\makebox(0,0)[b]{\smash{1}}}%
    \put(0.16375298,0.94875556){\color[rgb]{0,0,0}\makebox(0,0)[b]{\smash{$I_{1/2}$}}}%
    \put(0.50206142,0.94875556){\color[rgb]{0,0,0}\makebox(0,0)[b]{\smash{$I_{1/2}$}}}%
    \put(0.3627581,0.77960135){\color[rgb]{0,0,0}\makebox(0,0)[b]{\smash{$I_{2/2}$}}}%
    \put(0.66126555,0.77960135){\color[rgb]{0,0,0}\makebox(0,0)[b]{\smash{$I_{2/2}$}}}%
    \put(0.17065178,0.55561173){\color[rgb]{0,0,0}\makebox(0,0)[b]{\smash{$I_1$}}}%
    \put(0.35970643,0.50280903){\color[rgb]{0,0,0}\makebox(0,0)[b]{\smash{$I_2$}}}%
    \put(0.78066847,0.36169091){\color[rgb]{0,0,0}\makebox(0,0)[b]{\smash{$I_{N/2}$}}}%
    \put(0.56176295,0.36169091){\color[rgb]{0,0,0}\makebox(0,0)[b]{\smash{$I_{N/2}$}}}%
    \put(0.50896022,0.17750239){\color[rgb]{0,0,0}\makebox(0,0)[b]{\smash{$I_N$}}}%
  \end{picture}%
\endgroup%

%% file: figures/figures/LUMPEDcirctransmline.pdf_tex
%% Creator: Inkscape 0.48.3.1, www.inkscape.org
%% Accompanies image file 'LUMPEDcirctransmline.pdf' (pdf, eps, ps)
\begingroup%
  \makeatletter%
  \providecommand\color[2][]{%
    \errmessage{(Inkscape) Color is used for the text in Inkscape, but the package 'color.sty' is not loaded}%
    \renewcommand\color[2][]{}%
  }%
  \providecommand\transparent[1]{%
    \errmessage{(Inkscape) Transparency is used (non-zero) for the text in Inkscape, but the package 'transparent.sty' is not loaded}%
    \renewcommand\transparent[1]{}%
  }%
  \providecommand\rotatebox[2]{#2}%
  \ifx\svgwidth\undefined%
    \setlength{\unitlength}{283.46455078bp}%
    \ifx\svgscale\undefined%
      \relax%
    \else%
      \setlength{\unitlength}{\unitlength * \real{\svgscale}}%
    \fi%
  \else%
    \setlength{\unitlength}{\svgwidth}%
  \fi%
  \global\let\svgwidth\undefined%
  \global\let\svgscale\undefined%
  \makeatother%
  \begin{picture}(1,0.60000002)%
    \put(0,0){\includegraphics[width=\unitlength]{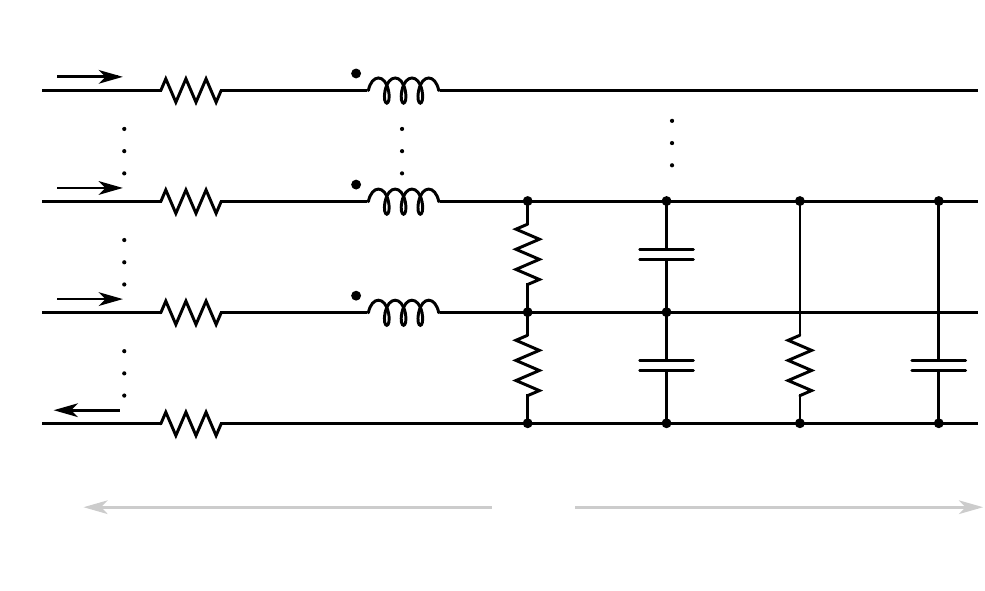}}%
    \put(0.01690852,0.14078857){\color[rgb]{0,0,0}\makebox(0,0)[b]{\smash{0}}}%
    \put(0.01690852,0.51896637){\color[rgb]{0,0,0}\makebox(0,0)[b]{\smash{N}}}%
    \put(0.10450452,0.53842304){\color[rgb]{0,0,0}\makebox(0,0)[b]{\smash{$i_{N}$}}}%
    \put(0.10450452,0.42553415){\color[rgb]{0,0,0}\makebox(0,0)[b]{\smash{$i_{j}$}}}%
    \put(0.10450452,0.31264525){\color[rgb]{0,0,0}\makebox(0,0)[b]{\smash{$i_{i}$}}}%
    \put(0.11586209,0.11323164){\color[rgb]{0,0,0}\makebox(0,0)[b]{\smash{$\sum_{j=1}^N i_j$}}}%
    \put(0.1888983,0.19374219){\color[rgb]{0,0,0}\makebox(0,0)[b]{\smash{$r_0\Delta z$}}}%
    \put(0.1888983,0.30663106){\color[rgb]{0,0,0}\makebox(0,0)[b]{\smash{$r_i\Delta z$}}}%
    \put(0.1888983,0.41951996){\color[rgb]{0,0,0}\makebox(0,0)[b]{\smash{$r_j\Delta z$}}}%
    \put(0.1888983,0.53240885){\color[rgb]{0,0,0}\makebox(0,0)[b]{\smash{$r_N\Delta z$}}}%
    \put(0.47112054,0.53240885){\color[rgb]{0,0,0}\makebox(0,0)[b]{\smash{$L_{NN}\Delta z$}}}%
    \put(0.47112054,0.41951996){\color[rgb]{0,0,0}\makebox(0,0)[b]{\smash{$L_{jj}\Delta z$}}}%
    \put(0.47112054,0.30663106){\color[rgb]{0,0,0}\makebox(0,0)[b]{\smash{$L_{ii}\Delta z$}}}%
    \put(1.03235264,0.21837648){\color[rgb]{0,0,0}\makebox(0,0)[b]{\smash{$C_{jj}\Delta z$}}}%
    \put(0.87424677,0.21837648){\color[rgb]{0,0,0}\makebox(0,0)[b]{\smash{$G_{jj}\Delta z$}}}%
    \put(0.7501304,0.21837648){\color[rgb]{0,0,0}\makebox(0,0)[b]{\smash{$C_{ii}\Delta z$}}}%
    \put(0.59766898,0.21837648){\color[rgb]{0,0,0}\makebox(0,0)[b]{\smash{$G_{ii}\Delta z$}}}%
    \put(0.60331342,0.33126537){\color[rgb]{0,0,0}\makebox(0,0)[b]{\smash{$G_{ij}\Delta z$}}}%
    \put(0.75571343,0.33126537){\color[rgb]{0,0,0}\makebox(0,0)[b]{\smash{$C_{ij}\Delta z$}}}%
    \put(0.54207728,0.07246991){\color[rgb]{0,0,0}\makebox(0,0)[b]{\smash{$\Delta z$}}}%
  \end{picture}%
\endgroup%

%% file: figures/figures/termincirc1.pdf_tex
%% Creator: Inkscape 0.48+devel, www.inkscape.org
%% Accompanies image file 'termincirc1.pdf' (pdf, eps, ps)
\begingroup%
  \makeatletter%
  \providecommand\color[2][]{%
    \errmessage{(Inkscape) Color is used for the text in Inkscape, but the package 'color.sty' is not loaded}%
    \renewcommand\color[2][]{}%
  }%
  \providecommand\transparent[1]{%
    \errmessage{(Inkscape) Transparency is used (non-zero) for the text in Inkscape, but the package 'transparent.sty' is not loaded}%
    \renewcommand\transparent[1]{}%
  }%
  \providecommand\rotatebox[2]{#2}%
  \ifx\svgwidth\undefined%
    \setlength{\unitlength}{120.40228271bp}%
    \ifx\svgscale\undefined%
      \relax%
    \else%
      \setlength{\unitlength}{\unitlength * \real{\svgscale}}%
    \fi%
  \else%
    \setlength{\unitlength}{\svgwidth}%
  \fi%
  \global\let\svgwidth\undefined%
  \global\let\svgscale\undefined%
  \makeatother%
  \begin{picture}(1,0.61550664)%
    \put(0,0){\includegraphics[width=\unitlength,page=1]{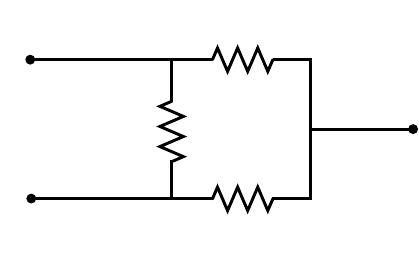}}%
    \put(0.06763522,0.50327532){\color[rgb]{0,0,0}\makebox(0,0)[b]{\smash{$1$}}}%
    \put(0.06763522,0.18434453){\color[rgb]{0,0,0}\makebox(0,0)[b]{\smash{$2$}}}%
    \put(0.2669671,0.29065487){\color[rgb]{0,0,0}\makebox(0,0)[b]{\smash{$R_{12}$}}}%
    \put(0.58236216,0.54935881){\color[rgb]{0,0,0}\makebox(0,0)[b]{\smash{$R_{11}$}}}%
    \put(0.58236216,0.01780742){\color[rgb]{0,0,0}\makebox(0,0)[b]{\smash{$R_{22}$}}}%
    \put(0.93307303,0.34855099){\color[rgb]{0,0,0}\makebox(0,0)[b]{\smash{O}}}%
  \end{picture}%
\endgroup%

%% file: figures/figures/termincirc2.pdf_tex
%% Creator: Inkscape 0.48+devel, www.inkscape.org
%% Accompanies image file 'termincirc2.pdf' (pdf, eps, ps)
\begingroup%
  \makeatletter%
  \providecommand\color[2][]{%
    \errmessage{(Inkscape) Color is used for the text in Inkscape, but the package 'color.sty' is not loaded}%
    \renewcommand\color[2][]{}%
  }%
  \providecommand\transparent[1]{%
    \errmessage{(Inkscape) Transparency is used (non-zero) for the text in Inkscape, but the package 'transparent.sty' is not loaded}%
    \renewcommand\transparent[1]{}%
  }%
  \providecommand\rotatebox[2]{#2}%
  \ifx\svgwidth\undefined%
    \setlength{\unitlength}{139.16341553bp}%
    \ifx\svgscale\undefined%
      \relax%
    \else%
      \setlength{\unitlength}{\unitlength * \real{\svgscale}}%
    \fi%
  \else%
    \setlength{\unitlength}{\svgwidth}%
  \fi%
  \global\let\svgwidth\undefined%
  \global\let\svgscale\undefined%
  \makeatother%
  \begin{picture}(1,0.53252788)%
    \put(0,0){\includegraphics[width=\unitlength,page=1]{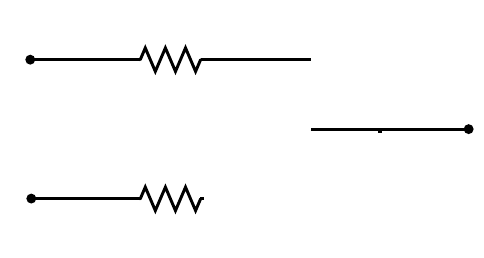}}%
    \put(0.05851707,0.43542685){\color[rgb]{0,0,0}\makebox(0,0)[b]{\smash{$1$}}}%
    \put(0.05851707,0.15949228){\color[rgb]{0,0,0}\makebox(0,0)[b]{\smash{$2$}}}%
    \put(0.81814091,0.32952572){\color[rgb]{0,0,0}\makebox(0,0)[b]{\smash{$R_{O}$}}}%
    \put(0.35438716,0.47529768){\color[rgb]{0,0,0}\makebox(0,0)[b]{\smash{$R_{1}$}}}%
    \put(0.35438716,0.01540677){\color[rgb]{0,0,0}\makebox(0,0)[b]{\smash{$R_{2}$}}}%
    \put(0.97479865,0.17353662){\color[rgb]{0,0,0}\makebox(0,0)[b]{\smash{O}}}%
    \put(0,0){\includegraphics[width=\unitlength,page=2]{termincirc2.pdf}}%
  \end{picture}%
\endgroup%

%% file: Conclusions.tex
\section{Conclusions}
\label{sec:conclusion}

In the case of signal formation in a long length multi conductor
detector, the corresponding current sources of the conductors, are
calculated with the same procedure as for a short multi conductor
detector. The difference is that the internal equivalent
(electrotechnical) circuit of the long detector is a system of
coupled transmission lines, unlike the short detector case where the
various capacitances of the electrodes are involved. If the
excitation (e.g. a passing particle) occurs far away from the ends
of the long detector, then the corresponding current sources are
connected to two identical multi conductor transmission lines, in
two opposite directions. The total auxiliary currents are split into
two sets of equal currents each one set feeding the corresponding
coupled transmission lines. What will happen to the propagating
signals in the two opposite directions after they reach the ends of
the lines depends on the loads at the terminals of the lines. If the
particle passes in a place very near the one end of the detector,
assuming this end to be "open" with no any external circuit attached
to it, and ignoring edge effects, then the current sources are
connected only  to  one multi conductor transmission line, the other
set of transmission lines mentioned above is a kind of an open
circuit. Then the set of currents feeds one set of transmission
lines (propagation along one direction), there is not the 1/2 in the
formula for the currents since there is not any splitting in this
case. The case of the cylindrical detector with only one wire
inside, is a simple case where the current source "sees" the
characteristic impedance of the line, which for an ideal (without
any type of lossless) line, is an ohmic resistance. We treated the
ideal case with no losses but one could guess that, for small
losses, the current sources will be the same as in the lossless case
while the propagation along  the transmission line will be treated
as in the lines with small losses. The problem of proper termination
of a multi conductor detector is a complicated problem. It is the
same problem for  multi conductor transmission line. In this work it
is proved (justified) that the problem of signals in long detectors
is split into two parts: a) the auxiliary currents from the passage
of a particle is calculated the same way it is done for small size
detectors, and b) the signals propagation follows the known
procedure of the multi conductor transmission lines in all respects,
including possible reflections.

%% file: Acknowledgments.tex
\section*{Acknowledgments}

I would like to thank Professor Theo Alexopoulos, because without
him, my thoughts about the subject of signal formation in multi
conductor detectors would be remained incomplete. He has, for many
years, a continuous involvement with Micromegas detectors and strong
will to understand as much as possible from the various aspects of
this type of detectors. As a result he  pressured me to examine in a
more detail the subject.
I thank Dr Stefanos Dris for making corrections and Mr Sotirios Fragkiskos for his help. \\

%\noindent The present work was co-funded by the European Union (European Social Fund ESF) and Greek national funds through the Operational Program "Education and Lifelong Learning" of the National Strategic Reference Framework (NSRF) 2007-1013. ARISTEIA-1893-ATLAS MICROMEGAS.